\documentclass[fleqn,usenatbib]{mnras}

\usepackage{newtxtext,newtxmath}

\usepackage[T1]{fontenc}

\DeclareRobustCommand{\VAN}[3]{#2}
\let\VANthebibliography\thebibliography
\def\thebibliography{\DeclareRobustCommand{\VAN}[3]{##3}\VANthebibliography}

\usepackage{graphicx}	
\usepackage{amsmath}	
\usepackage{pdflscape}

\title[The molecular clouds of massive galaxies]{Two distinct molecular cloud populations detected in massive galaxies}

\author[Tom Rose et al.]{
Tom Rose$^{1,2}$\thanks{E-mail: thomas.rose@uwaterloo.ca},
B. R. McNamara$^{1,2}$,
F. Combes$^{3}$,
A. C. Edge$^{4}$,
M. McDonald$^{5}$,\newauthor
Ewan O'Sullivan$^{6}$,
H. Russell$^{7}$,
A. C. Fabian$^{8}$,
G. Ferland$^{9}$,
P. Salom\'e$^{3}$,
G. Tremblay$^{6}$,
\\
$^{1}$Department of Physics and Astronomy, University of Waterloo, Waterloo, ON N2L 3G1, Canada \\
$^{2}$Waterloo Centre for Astrophysics, Waterloo, ON N2L 3G1, Canada \\
$^{3}$LERMA, Observatoire de Paris, PSL Univ., College de France, CNRS, Sorbonne Univ., Paris, France\\
$^{4}$Centre for Extragalactic Astronomy, Durham University, DH1 3LE, UK\\
$^{5}$MIT Kavli Institute, 70 Vassar Street, Cambridge, MA 02139, USA \\
$^{6}$Center for Astrophysics $|$ Harvard \& Smithsonian, 60 Garden St., Cambridge, MA 01238, USA\\
$^{7}$School of Physics \& Astronomy, University of Nottingham, Nottingham, NG7 2RD, UK\\
$^{8}$Institute of Astronomy, Cambridge University, Madingley Rd., Cambridge, CB3 0HA, UK\\
$^{9}$Department of Physics and Astronomy, University of Kentucky, Lexington, Kentucky 40506-0055, USA\\
}

\date{Accepted XXX. Received YYY; in original form ZZZ}

\pubyear{2024}

\begin{document}
\label{firstpage}
\pagerange{\pageref{firstpage}--\pageref{lastpage}}
\maketitle

\begin{abstract}
We present new ALMA observations of CO, CN, CS, HCN and HCO$^{+}$ absorption seen against the bright and compact radio continuum sources of eight massive galaxies. Combined with archival observations, they reveal two distinct populations of molecular clouds, which we identify by combining CO emission and absorption profiles to unambiguously reveal each cloud's direction of motion and likely location. In galaxy disks, we see clouds with low velocity dispersions, low line of sight velocities and a lack of any systemic inflow or outflow. In galactic cores, we find high velocity dispersion clouds inflowing at up to 550\,km/s. This provides observational evidence in favour of cold accretion onto galactic centres, which likely contributes to the fuelling of active galactic nuclei. We also see a wide range in the \mbox{CO(2-1)/CO(1-0)} ratios of the absorption lines. This is likely the combined effect of hierarchical substructure within the molecular clouds and continuum sources which vary in size with frequency.
\end{abstract}

\begin{keywords}
galaxies: active -- galaxies: clusters: general -- radio continuum: galaxies -- radio lines: ISM
\end{keywords}



\section{Introduction}

\begin{table*}
    \caption{The ALMA observations presented in this paper.}
    \centering
    \begin{tabular}{lcccccccr}
    \hline
    Source & Ang. res. & FOV & Int. Time & Obs. date & Line(s) & Cont. Flux Density & Project code \\
     & (") & (") & (s) & (yyyy-mm-dd) & & (mJy, GHz) & \\
    \hline
    NGC 6868 & 0.81 & 58.8 & 5100 & 2018-01-25 & CO(1-0) & 41, 107 & 2017.1.00629 \\
    NGC 6868 & 0.11 & 25 & 1400 & 2021-11-20 & CO(2-1), CN(2-1) & 33, 234 &  2021.1.00766 \\
    NGC 6868 & 0.16 & 34.4 & 2800 & 2021-11-19 & HCO$^{+}$(2-1), HCN(2-1) & 36, 169 &  2021.1.00766 \\
    Abell S555 & 0.63 & 56 & 2800 & 2018-01-23 & CO(1-0) & 32, 103 & 2017.1.00629 \\
    Abell S555 & 0.14 & 25.8 & 3700 & 2022-07-20 & CO(2-1), CN(2-1) & 23, 226 & 2021.1.00766 \\
    Abell S555 & 0.26 & 33.2 & 1800 & 2022-07-05 & HCO$^{+}$(2-1), HCN(2-1) & 25, 175 & 2021.1.00766 \\
    Hydra-A & 2.07 & 57 & 2700 & 2018-07-18 & CO(1-0) & 99, 103 & 2017.1.00629 \\
    Hydra-A & 0.26 & 26 & 5700 & 2018-10-30 & CO(2-1), CN(2-1) & 60, 223 & 2018.1.01471 \\
    Hydra-A & 0.34 & 33.5 & 5100 & 2018-11-16 & HCO$^{+}$(2-1), HCN(2-1) & 95, 168 &  2018.1.01471 \\
    Abell 2390 & 0.40 & 59.0 & 8000 & 2018-01-07 & CO(1-0) & 22, 98 & 2017.1.00629 \\
    Abell 2390 & 0.58 & 30.4 & 4800 & 2022-05-20 & CO(2-1) & 11, 192 & 2021.1.00766 \\
    Abell 2390 & 0.21 & 42.1 & 900 & 2021-11-22 & HCO$^{+}$(2-1), HCN(2-1) & 14, 138 & 2021.1.00766 \\
    RXCJ0439.0+0520 & 0.18 & 29.8 & 2600 & 2022-07-22 & CO(2-1), CN(2-1) & 72, 195 & 2021.1.00766 \\
    RXCJ0439.0+0520 & 0.30 & 41.4 & 1300 & 2021-12-04 & HCO$^{+}$(2-1), HCN(2-1) & 85 ,141 & 2021.1.00766 \\
    Abell 1644 & 1.74 & 57 & 2800 & 2018-08-21 & CO(1-0) & 67, 103 & 2017.1.00629 \\
    Abell 1644 & 0.47 & 25.9 & 1300 & 2022-06-08 & CO(2-1), CN(2-1) & 38, 225 & 2021.1.00766 \\
    Abell 1644 & 0.19 & 33.3 & 1700 & 2022-07-22 & HCO$^{+}$(2-1), HCN(2-1) & 39, 175 & 2021.1.00766 \\
    Circinus & 0.027 & 16.7 & 5600 & 2019-07-14 & CO(3-2), HCO$^{+}$(4-3), HCN(4-3) & 37, 259 & 2018.1.00581 \\
    Circinus & 0.020 & 22.5 & 5400 & 2019-06-07 & HCO$^{+}$(3-2) & 29, 350 & 2018.1.00581\\
    IC 4296 & 0.40 & 25 & 1700 & 2014-07-23 & CO(2-1) & 190, 226 & 2013.1.00229 \\
    Abell 2597 & 0.43 & 26 & 10900 & 2013-11-17 & CO(2-1) & 15, 213 & 2012.1.00988 \\
    Abell 2597 & 0.17 & 34.2 & 2700 & 2021-11-25 & HCO$^{+}$(2-1), HCN(2-1) & 14, 170 & 2021.1.00766 \\
    NGC 5044 & 0.74 & 54 & 2400 & 2018-09-20 & CO(1-0) & 32, 107 & 2017.1.00629 \\
    NGC 5044 & 1.89 & 25 & 1400 & 2012-01-13 & CO(2-1) & 45, 228 & 2011.0.00735 \\
    NGC 5044 & 0.18 & 34.4 & 2600 & 2022-07-22 & HCO$^{+}$(2-1), HCN(2-1) &  38, 170 & 2021.1.00766 \\
    NGC 1052 & 0.48 & 54.0 & 1200 & 2016-05-26 & CO(1-0) & 794, 108 & 2015.1.00591 \\
    NGC 1052 & 0.03 & 26.2 & 600 & 2015-10-23 & CO(2-1) & 443, 223 & 2015.1.01290 \\
    NGC 1052 & 0.13 & 16.8 & 4300 & 2015-08-16 & CO(3-2) & 333, 344 & 2013.1.01225 \\
    NGC 1052 & 0.26 & 62.8 & 700 & 2017-07-28 & HCN(1-0) & 1020, 88 & 2016.1.00375 \\
    NGC 4261 & 0.26 & 24.7 & 1900 & 2018-01-19 & CO(2-1) & 253, 236 & 2017.1.00301 \\
    NGC 4261 & 0.18 & 16.8 & 2300 & 2018-01-21 & CO(3-2), HCO$^{+}$(4-3), HCN(4-3) & 223, 348 & 2017.1.01638 \\
    \hline
    \end{tabular}
    \label{tab:observations}
\end{table*}

Our understanding of the molecular gas in massive galaxies has been built upon theory, simulations, and observations \citep[examples include][]{ODea1994, Pizzolato2005, Gaspari2010, McNamara2016, Gaspari2018, Olivares2019, Rose2019b, Fabian2022}. Although theories and simulations can be readily used to study molecular gas on a wide range of spatial scales, observations are more limited. In extragalactic sources, molecular gas is most readily observed in emission, but this typically limits us to scales of $10^{5}\textnormal{M}_{\odot}$ or more. They are therefore orders of magnitude away from resolving cold gas into individual clouds -- the scales on which key processes such as star formation and AGN accretion occur.

An alternative technique to study molecular gas is via absorption lines against a bright and compact background continuum source. This technique is advantageous because its sensitivity is many orders or magnitude higher than emission. As a result, the molecular gas can be studied on much smaller scales. The iconic radio galaxy Centaurus-A was the first system to be studied in this way \citep{Israel1990}. Of particular advantage is that this system's radio continuum emission and absorbing clouds lie within the same galaxy, meaning that any redshifted absorption unambiguously indicates movement towards the galaxy centre along the line of sight. This is in contrast to both emission lines and extrinsic absorption lines \citep[i.e. absorption found due to the chance alignment of a background quasar and an intervening galaxy, e.g.][]{Wiklind1996a, Wiklind1997b}. In these cases, there is ambiguity as to whether the observed gas lies in front of or behind the core of the galaxy, so its direction of motion is unclear.

Unfortunately, finding molecular absorption like that in Centaurus-A requires an extremely bright and compact radio continuum source, so a lack of similar observations followed over the next two decades. However, with the much improved angular resolution and sensitivity offered by the Atacama Large Millimeter/submillimeter Array (ALMA), the last decade or so has been more fruitful and many more intrinsic absorption line systems have now been found, mostly in group and cluster environments \citep[e.g.][]{David2014, Tremblay2016, Rose2019a, Rose2019b, Ruffa2019, Rose2020, Kameno2020}. With these advancements, individual cases of molecular and atomic absorption have now been suggested as being due to clouds within the circumnuclear disk and/or accreting onto the central supermassive black hole \citep[e.g.][]{Tremblay2016, Rose2019b, Morganti2023, Izumi2023, Oosterloo2023}.

\citet{Rose2019b} presented the first multi-target study of intrinsic molecular absorption with ALMA observations of CO(1-0) and CN(1-0). Added to Centaurus-A, NGC 5044 and Abell 2597, this raised the total number of intrinsic molecular absorption line systems to eight. Initial results from this survey appeared to show that molecular absorption in such galaxies is surprisingly common, with a detection rate of around 40 per cent. Additionally, although most of the absorbing gas clouds had no significant line of sight velocity towards or away from their galaxy centre, a slight bias for clouds to be moving towards their galaxy's centre appeared to be present.

Subsequent observations of the edge-on elliptical brightest cluster galaxy Hydra-A have shown that higher order transitions and a wider mix of molecular tracers can be particularly effective at detecting intrinsic molecular absorption \citep{Rose2020}. As such, in this paper we present new ALMA observations of eight massive galaxies covering CO(2-1), CN(2-1), HCO$^{+}(2-1)$ and HCN(2-1) in many systems with molecular absorption already detected via other lines. These new observations offer a significant improvement -- detecting many more molecular absorption regions and with a higher sensitivity. Combined with CO(2-1) observations, we can therefore study the bulk motions of the absorbing clouds in the context of each galaxy's molecular emission. We try to be exhaustive in studying all known intrinsically absorbing systems. Although a handful of additional intrinsic molecular absorption systems are known, they are without high resolution observations of the molecular emission. These include \citet{Allison2019}, \citet{Combes2024} and \citet{Emonts2024}.

The rest of this paper is laid out as follows. In section 2 we give details on our sample, and present moments maps and spectra. We also detail our data reduction methods. In section 3 we calculate excitation temperatures and column densities for the absorbing gas in each system, as well as molecular masses from the emission. In section 4, we discuss our findings and the potential implications. Section 5 summarises our findings. We assume $\Lambda$ cold dark matter cosmology with $\Omega_{\textnormal{M}} =0.3$ and $\Omega_{\Lambda} = 0.7$, and a Hubble constant of H$_{0} = 70$\,km/s/Mpc.

\section{Data}

\begin{table*}
\centering
	\begin{tabular}{lcccccc}
\hline
Source & RA, Dec (J2000) & $z_{*}$ & Redshift reference & $v_{*}$ & $v_{\textnormal{mol}} - v_{*}$ \\
\hline
  NGC 6868 & 20:09:54.1, -48:22:46.3 & 0.0095$\pm$0.0001 (FORS) & \citet{Rose2019b} &  2830$\pm$30 & -60$\pm30$ \\
  Abell S555 & 05:57:12.5, -37:28:36.5 & 0.0446$\pm$0.0001 (MUSE) & \citet{Rose2019b} & 13364$\pm$30 & -190$\pm40$ \\
  Hydra-A & 09:18:05.7, -12:05:44.0 & 0.0544$\pm$0.0001 (MUSE) & \citet{Rose2019b} &  16294$\pm$30 & 120$\pm50$ \\
  Abell 2390 & 21:53:36.8, +17:41:43.7 & 0.2304$\pm$0.0001 (VIMOS) & \citet{Hamer2016} & 69074$\pm$30 & 170$\pm40$ \\
  RXCJ0439.0+0520 & 04:39:02.3, +05:20:43.7 & 0.2076$\pm$0.0001 (VIMOS) & \citet{Hamer2016} &  62237$\pm30$ & 210$\pm60$ \\
  Abell 1644 & 12:57:11.6, -17:24:34.1 & 0.0473$\pm$0.0001 (MUSE) & \citet{Rose2019b} & 14191$\pm$30 & -60$\pm40$ \\
  Circinus & 14:13:09.9, -65:20:21.0 & 0.00145$\pm$0.00001 (NED) & \citet{Ursini2023} &  $434\pm3$ & - & \\
  IC 4296 & 13:36:39.1, -33:57:57.3 & 0.01247$\pm$0.00003 (NED) & \citet{Ruffa2019} &  3738$\pm$10 & -60$\pm50$ \\
  Abell 2597 & 23:25:19.7 , -12:07:27.7 & 0.0821$\pm$0.0001 (MUSE) & \citet{Tremblay2018} & 24613$\pm$30 & 20$\pm40$ \\
  NGC 5044 & 13:15:24.0, -16:23:07.6 & 0.0092$\pm$0.0001 (MUSE) & \citet{Rose2019b} &  2761$\pm$30 & 30$\pm40$ & \\
  NGC 1052 & 02:41:04.8, -08:15:20.8 & 0.00498$\pm$0.00002 (RC3) & \citet{deVauc1991} &  $1492\pm5$ & - \\
  NGC 4261 & 12:19:23.2, +05:49:29.7 & 0.0073$\pm$0.0001 (OHP) & \citet{Matsumoto2001} &  2200$\pm$30 & -10$\pm40$ \\
  Centaurus-A & 13:25:27.6, -43:01:09.0 & $0.00183\pm0.00002$ (NED) & \citet{Crook2007} & 547$\pm$5 & - \\
		\hline 
	\end{tabular}
\caption{Details on the position and redshift of each source. All redshifts are barycentric and use the optical convention. RC3 is the Third Reference Catalogue of Bright Galaxies \citep{deVauc1991}. Velocities are in km s$^{-1}$.}

    \label{tab:gas_properties_and_redhsifts}
\end{table*}

\begin{table*}
    \centering
    \begin{tabular}{lcccccc}
    \hline
    Source & Galaxy Type & Inclination / $^{\circ}$ & Molecular Morphology & Molecular Mass / M$_{\odot}$ \\
    \hline
    NGC 6868 & Elliptical & 41 & 0.5\,kpc wide disk inclined at $\approx70^{\circ}$ & 1.3$ \pm 0.2\times 10^{8}$ \\
    Abell S555 & Unknown & - & Plume of 2.5\,kpc as measured along l.o.s. & 5.6$ \pm 0.5 \times 10^{8}$ \\
    Hydra-A & Elliptical-S0 & 79 & 5\,kpc molecular disc inclined at $\approx 90^{\circ}$ & 3.1$ \pm 0.2 \times 10^{9}$ \\
    Abell 2390 & Irregular & - & Plume of 15\,kpc as measured along l.o.s. & 2.2$ \pm 0.2 \times 10^{10}$ \\
    RXCJ0439.0+0520 & Irregular & - & Plume of 9\,kpc as measured along l.o.s. & 1.3$ \pm 0.4 \times 10^{10}$ \\
    Abell 1644 & Irregular & - & Clumps distributed along an arc & 1.6$ \pm 0.3 \times 10^{9}$  \\
    Circinus & Sb spiral & 64 & Spiral arms inclined at $\approx65^{\circ}$ \citep{Izumi2018} & - \\
    IC 4296 & Elliptical & 90 & 0.5\,kpc wide disk inclined at $\approx74^{\circ}$ & 2.3$ \pm 0.2 \times 10^{7}$ \\
    Abell 2597 & Elliptical & 68 & Irregular central 7\,kpc, faint emission to 30\,kpc & 4.6$ \pm 0.5 \times 10^{9}$ \\
    NGC 5044 & Elliptical & 43 & One or two weak clumps 200-500\,pc from AGN & 2.4$ \pm 0.4 \times 10^{7}$ \\
    NGC 1052 & Elliptical & 70 & Gas poor with $5.3\times10^{5}\textnormal{M}_{\odot}$ CND \citep{Kameno2020} & -\\
    NGC 4261 & Elliptical & 56 & 200\,pc wide disk inclined at $\approx42^{\circ}$ & 2.1$ \pm 0.2 \times 10^{7}$ \\
    Centaurus-A & S0 & 45 & Dense gas along dust lane \citep{Wild1997}  & $1.4\pm0.2\times10^{9*}$\\
    \hline
    \end{tabular}
    \caption{The classification of each galaxy and a brief summary of the molecular emission properties. Inclinations are taken from the HyperLeda database. For sources whose morphology and velocity distribution indicate a molecular disk, we estimate that disk's inclination using $\cos{i} = \sqrt{ [(b/a)^2 - r_{0}^{2}] / [1 - r_{0}^{2}]}$, where $a$ is the semi-major axis length, $b$ the semi-minor axis length and $r_{0}$ is the disk thickness \citep{Tully1998}. We also assume circular disk and $r_{0}^{2} = 0.2$, as suggested by \citet{Tully1998}. \\$^{*}$The molecular mass of Centaurus-A is taken from \citet{Parkin2012}.}
    \label{tab:molecular_properties}
\end{table*}

\subsection{Sample}

We present ALMA observations of molecular gas in 12 galaxies. Details of these observations are given in Table \ref{tab:observations}. Further information on the location and redshift of each source is given in Table \ref{tab:gas_properties_and_redhsifts}, while Table \ref{tab:molecular_properties} gives each galaxy's classification and summarises its molecular emission properties. More details on each source's properties are given in Appendix \ref{sec:Appendix}. 

The largest set of data we use is from ALMA project 2021.1.00766.S (P.I. Tom Rose), which aimed to detect CO(2-1), CN(2-1), HCO$^{+}$(2-1) and HCN(2-1) in several galaxies known to have molecular absorption from other lines. This project followed on from a survey of 23 X-ray selected brightest cluster galaxies (2017.1.00629.S, P.I. Alastair Edge) which aimed to detect CO(1-0) absorption against the galaxies' radio cores \citep[see][in which more details on the sample selection can be found]{Rose2019b}. In brief, by selecting the host cluster from its diffuse X-ray emission, this survey was not biased to central galaxies with any particular orientation of their AGN.

In this paper we aim to provide analysis of the bulk motions of all known intrinsic molecular absorption lines which have complementary high resolution molecular emission line observations, so data from a number of other projects is also presented. In Table \ref{tab:observations}, we show the project from which each observation originated. We do not include NGC 1275, which has low significance absorption reported by \citet{Nagai19}. Subsequent IRAM (Institut de Radioastronomie Millim\'{e}trique) observations with NOEMA (NOrthern Extended Millimeter Array) at higher sensitivity do not find molecular absorption in this system.

\subsection{Moments maps and spectra}

In Figures \ref{fig:NGC6868_maps_and_spectra} -- \ref{fig:IC4296_maps_and_spectra}, we show intensity, velocity, and velocity dispersion maps made with each galaxy's molecular emission. We indicate the location of each galaxy's compact and unresolved radio continuum source with a cross, against which we extract the spectra of CO(1-0), CO(2-1), CN(2-1), HCO$^{+}$(2-1) and HCN(2-1). Typically, the ALMA observations span velocities from around -1500 to +1500\,km/s, though we see no absorption or emission features outside the range shown in the spectra. Sometimes, absorption from other lines is also present. Abell 1644 has CS(5-4) absorption. Hydra-A has several more molecules detected in absorption than we show, including HCO$^{+}$(1-0) HCN(1-0), H$_{2}$C(3-2) and $^{13}$CO(2-1). These can be seen in \citep{Rose2020}, along with the best fits. NGC 1052 has several more lines shown in \citet{Kameno2020}. Finally, NGC 6868, Abell 2597, Abell 2390, Hydra-A and NGC 4261 have HI absorption. We clearly distinguish these HI spectra by showing them in green. In Figure \ref{fig:Centaurus-A_maps_and_spectra}, we also show HCO$^{+}$(2-1) and HCN(2-1) spectra seen against the central radio source of the iconic radio galaxy Centaurus-A, from \citet{Wiklind1997a}.

\subsection{Data reduction}

We produce data cubes from the ALMA observations using measurement sets available from the National Radio Astronomy Observatory (NRAO) Archive. We processed the data with \texttt{CASA} version 6.6.3.22, a software package produced and maintained by NRAO \citep{CASA}. From the measurement sets provided, we made continuum subtracted images using the CASA tasks \texttt{tclean} and \texttt{uvcontsub}. In \texttt{tclean}, we have used a hogbom deconvolver and produced images with the narrowest possible channel width. For our moments maps, we use a data cube made with natural weighting to minimise noise. In most cases, we extract our spectra from data cubes made with uniform weighting, which optimises the angular resolution of the data. This limits contamination from spatially extended molecular emission in the spectra. However, where no detectable emission is present close to the radio core, we extract spectra using a cube made with natural weighting.

The moments maps we show are produced from the CO(2-1) observations of each galaxy, except for Circinus and NGC 5044. In Circinus, we use very high angular resolution HCO$^{+}$(3-2) emission which traces the inner few parsecs of the galaxy. In NGC 5044, the CO(2-1) observation has a clear detection of emission, but is of very poor angular resolution. On the other hand, the CO(1-0) observation is less strong, but is of higher angular resolution. We therefore show an intensity map made from the CO(1-0) data, but velocity and velocity dispersion maps made from CO(2-1) data. 

We produce our maps using the \texttt{CASA} task \texttt{immoments}. Normally moment 0 maps are made using all channel values in each spaxel. However, for sources with a large beam and overlapping emission and absorption, we use `includepix = [0,100]', i.e. all pixel values from 0 to an arbitrarily large value. This prevents the negative absorption fluxes wiping out positive emission at other velocities. After producing these maps, we therefore estimate the summed value of the remaining positive noise in an emission free region, and subtract it from the whole image. This may bias the overall intensity map to be lower than reality because negative noise will effectively be subtracted from higher significance emission twice. However, this method more accurately displays the emission seen against the radio core. Additionally, this method only affects the moment 0 maps.

Our moments 1 and 2 maps are made with $>4\,\sigma$ emission, which we find to give the best appearance. To reduce noise in the data cubes, these maps have sometimes been made using a slightly smoothed image. The effect this has on the effective image resolution is indicated by the beam size on each image. This smoothing means we can capture the velocity and velocity dispersion of more extended emission.

\subsection{Line Fitting Procedure}

To determine the optimal fit for each spectrum, we first identify the minimum number of components required by incrementally adding Gaussian lines. We do this until no channel contains $>5\sigma$ emission and the RMS value of the fitting region is less than 1.5 times that of the remainder of the spectrum. This threshold provides a good fit while not introducing unnecessary or unwarranted complexity.

We then perform an initial fit using \texttt{Python}'s \texttt{SciPy} library. Specifically, we use the \texttt{curve fit} minimization algorithm. Our fitting procedure is similar for both emission and absorption lines. However, when a spectrum contains overlapping emission and absorption features, the absorption must be masked while we first fit to the emission. This fit to the emission is then subtracted from the spectrum before fitting to the absorption. We do this masking by eye, which we find to be effective. Additionally, we find that using a reasonable range of masking boundaries results in minimal changes to the ultimate fits.

After making a best fit to a spectrum, we find the noise level. We take this to be the root mean square (RMS) of the continuum emission, excluding regions exhibiting emission or absorption features. Then, 10,000 simulated spectra are generated based upon the observed spectrum and the initial best fit. At each velocity channel, a pseudo random value is drawn from a Gaussian distribution centred at 0 and with a dispersion equal to the RMS noise. This value, which may be positive or negative, is then added to the original spectrum's best fit value for that velocity channel. 

After repeating this for all velocity channels, a new simulated spectrum is produced. For each simulated spectrum, we then find a best fit in the same way as for the original spectrum. The 1$\sigma$ upper and lower error bounds are then defined by the values that delineate the 15.865 per cent highest and lowest results among the 10,000 fits. This ensures that 68.27 per cent of the fitted parameters fall within the 1\,$\sigma$ range. 

Where the emission and absorption overlap, we do not account for errors from fitting to the emission propagating through to the absorption fits. This is because the former are generally much smaller.

Best fit parameters for each spectrum will be made available online as supplementary material.

\begin{figure*}
    \includegraphics[width=\textwidth]{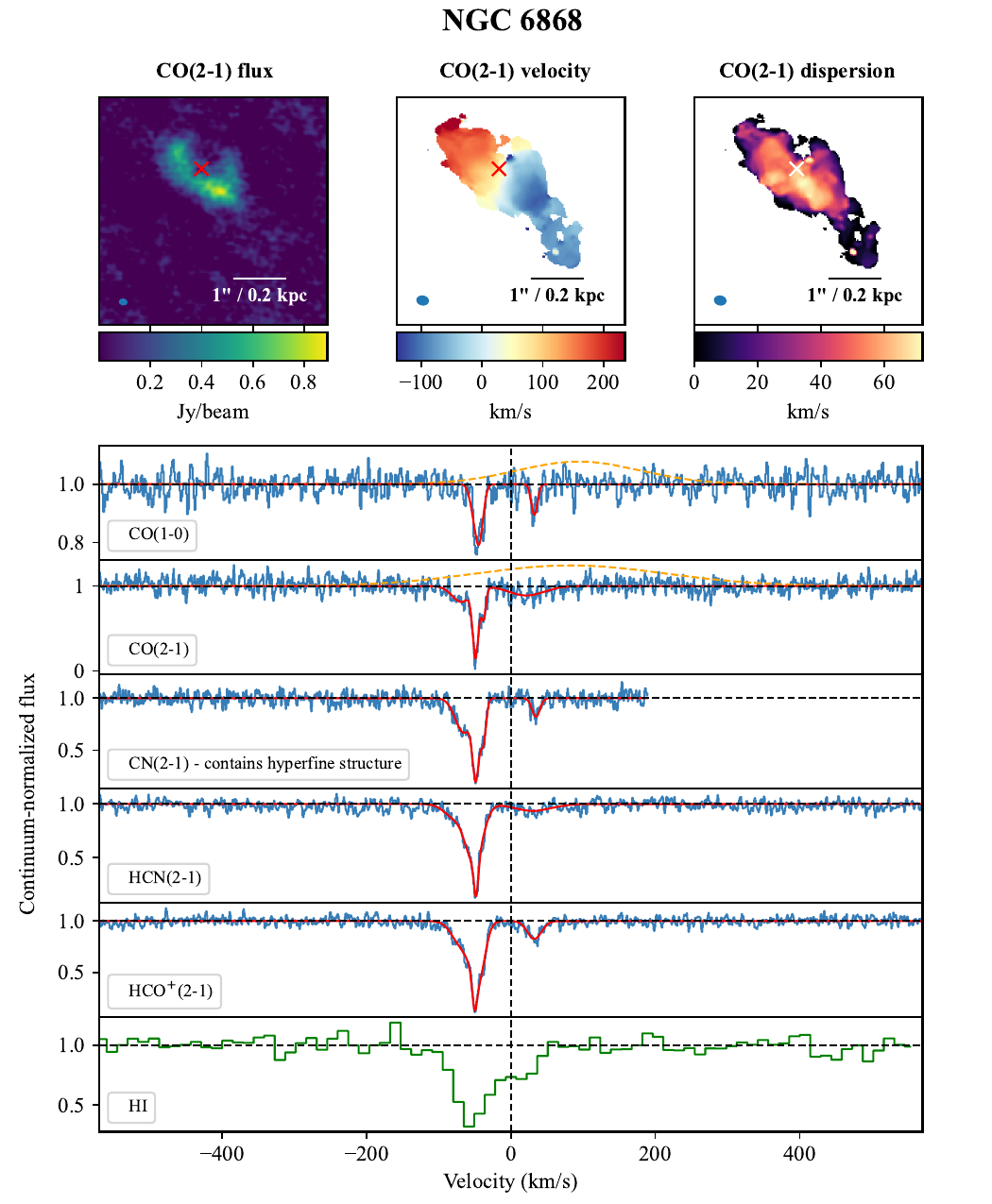}
    \caption{Intensity, velocity and velocity dispersion maps of NGC 6868 made from CO(2-1) emission. Crosses indicate the location of the radio continuum source, against which the spectra of several different molecular lines are extracted and shown below. An H\small{I} spectrum also extracted against the radio core is shown in green (Tom Oosterloo, private communication). Dashed yellow lines show CO emission subtracted from the spectra.}
    \label{fig:NGC6868_maps_and_spectra}
\end{figure*}

\begin{figure*}
	\includegraphics[width=\textwidth]{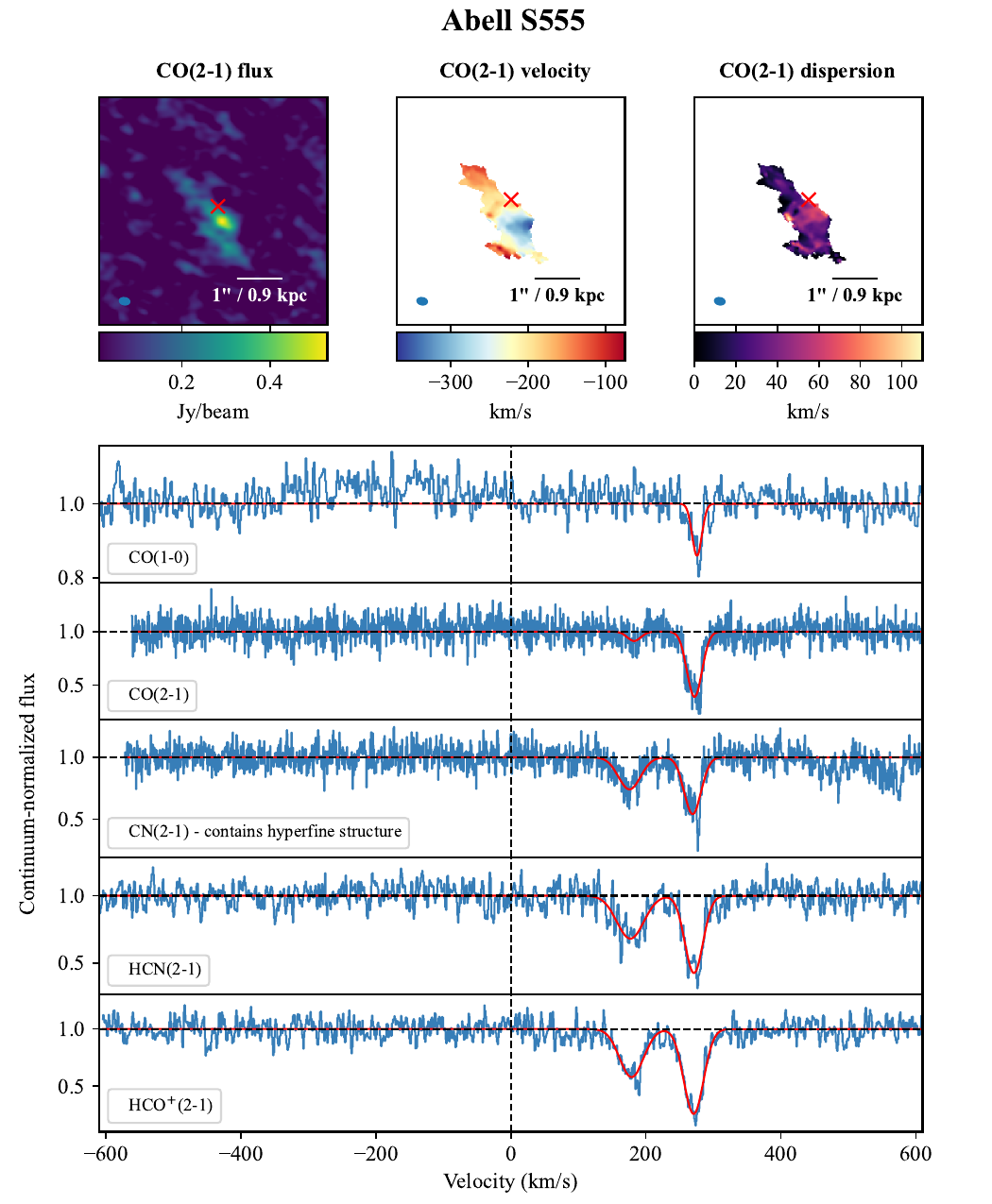}
    \caption{Intensity, velocity and velocity dispersion maps of Abell S555 made from CO(2-1) emission. Crosses indicate the location of the radio continuum source, against which the spectra of several different molecular lines are extracted and shown below. The CO(1-0) observation has a larger beam size than the other observations, so its spectrum contains some of the more extended emission.}
    \label{fig:S555_maps_and_spectra}
\end{figure*}

\begin{figure*}
	\includegraphics[width=0.99\textwidth]{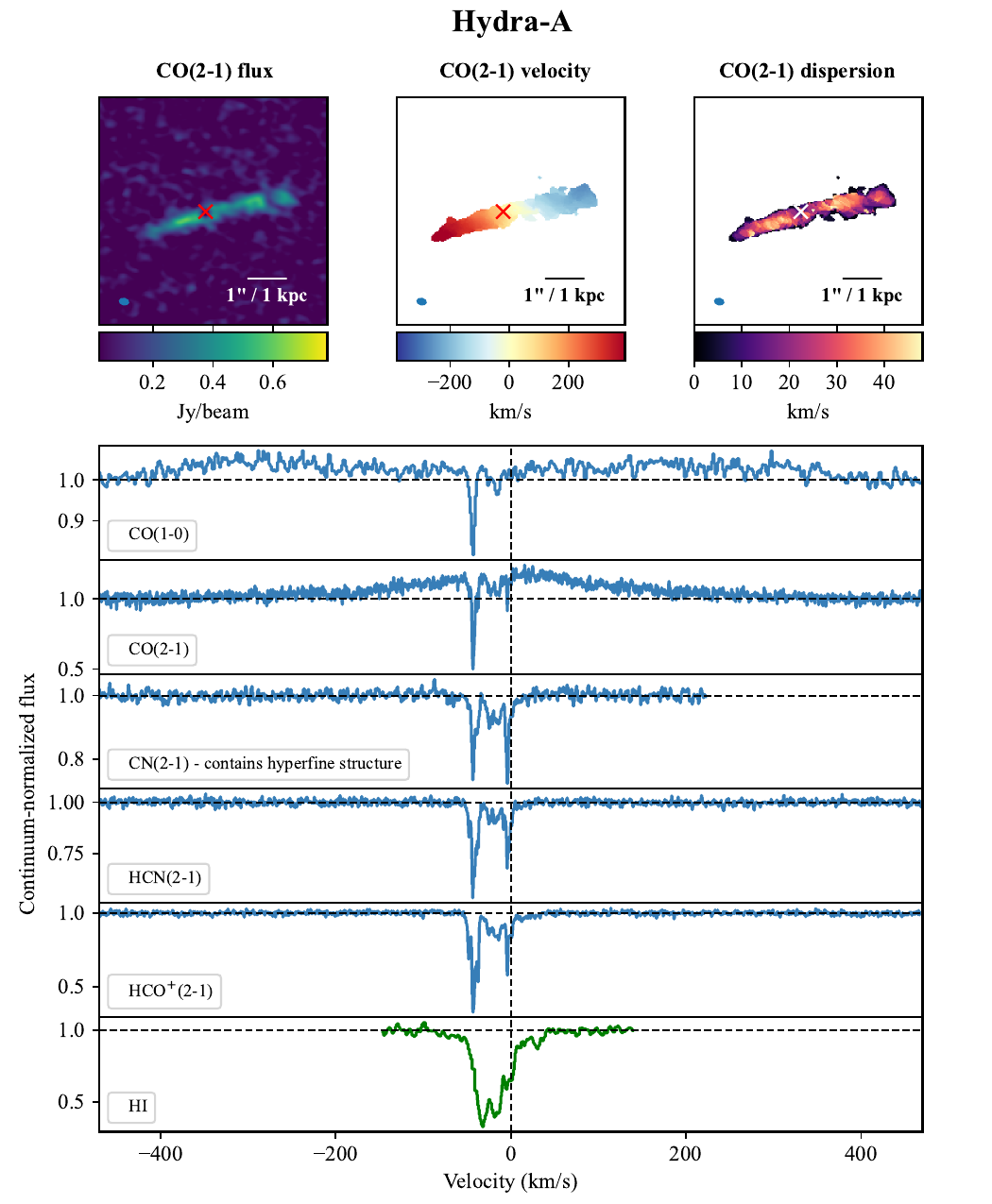}
    \caption{Intensity, velocity and velocity dispersion maps of Hydra-A made from CO(2-1) emission. Crosses indicate the location of the radio continuum source, against which the spectra of several different molecular lines are extracted and shown below. An H\small{I} spectrum also extracted against the radio core is shown in green \citep{Taylor1996}. Several more molecules are detected in absorption than presented here, including HCO$^{+}$(1-0) HCN(1-0), H$_{2}$C(3-2) and $^{13}$CO(2-1). Spectra of these lines are shown in \citep{Rose2020}. Due to the complex nature of the absorption, we do not show best fits in the spectra above, but they can also be seen in \citep{Rose2020}.}
    \label{fig:HydraA_maps_and_spectra}
\end{figure*}

\begin{figure*}
	\includegraphics[width=\textwidth]{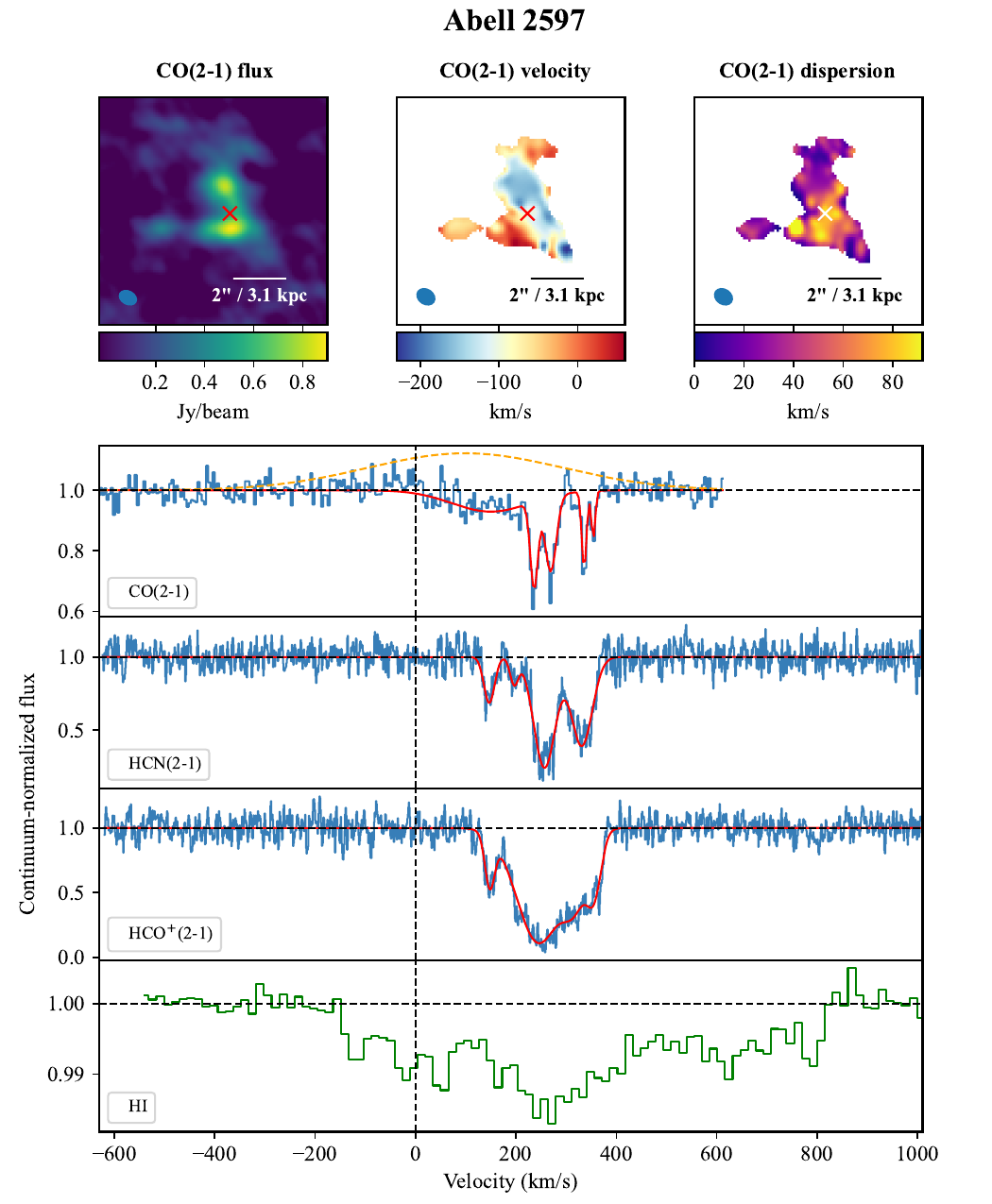}
    \caption{Intensity, velocity and velocity dispersion maps of Abell 2597 made from CO(2-1) emission. Crosses indicate the location of the radio continuum source, against which the spectra of several different molecular lines are extracted and shown below. An H\small{I} spectrum also extracted against the radio core is shown in green \citep{Hernandez2008}. The dashed yellow line shows subtracted CO emission.}
    \label{fig:A2597_maps_and_spectra}
\end{figure*}

\begin{figure*}
	\includegraphics[width=\textwidth]{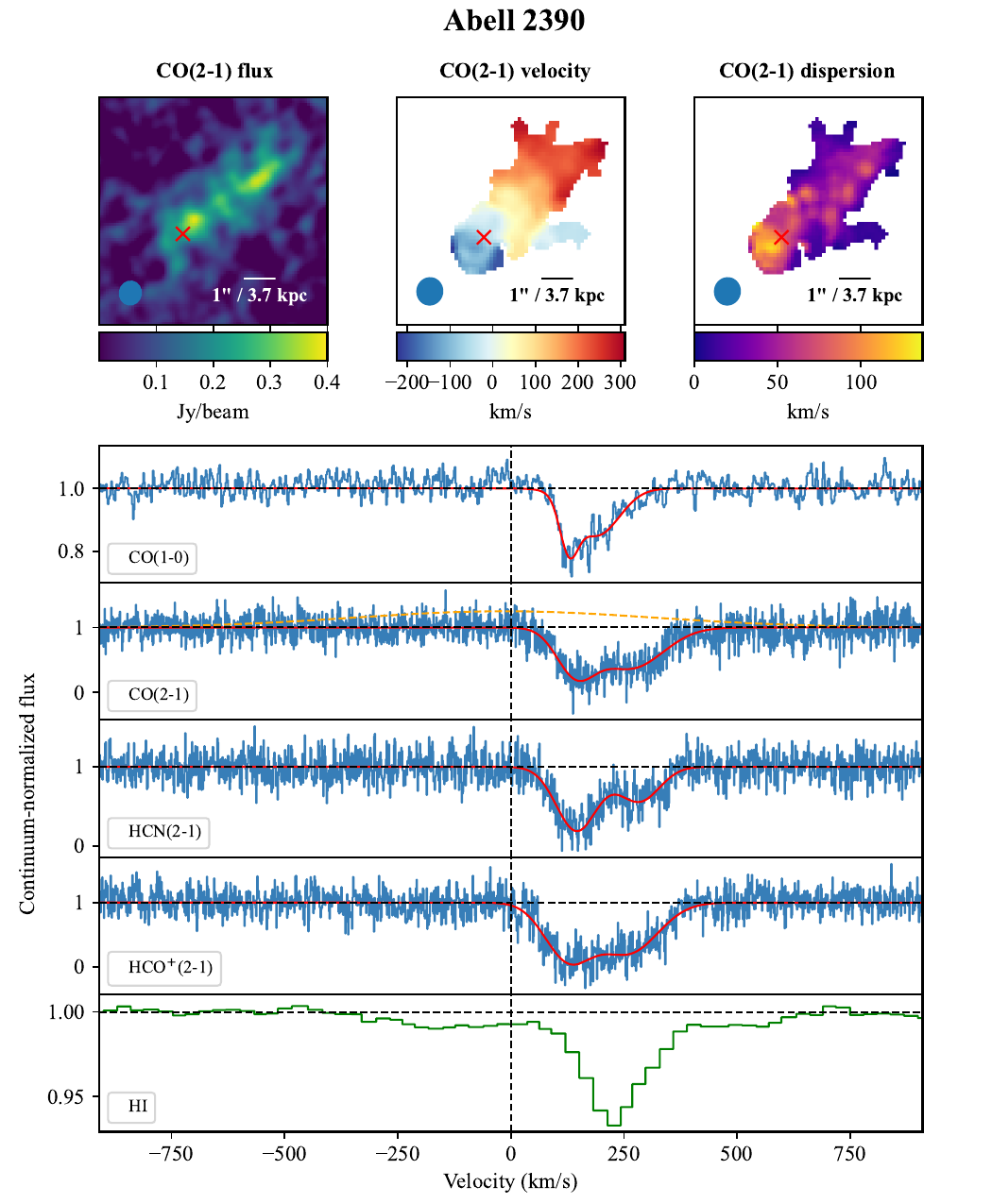}
    \caption{Intensity, velocity and velocity dispersion maps of Abell 2390 made from CO(2-1) emission. Crosses indicate the location of the radio continuum source, against which the spectra of several different molecular lines are extracted and shown below. An H\small{I} spectrum also extracted against the radio core is shown in green \citep{Hernandez2008}. The dashed yellow line shows subtracted CO emission.}
    \label{fig:A2390_maps_and_spectra}
\end{figure*}

\begin{figure*}
	\includegraphics[width=\textwidth]{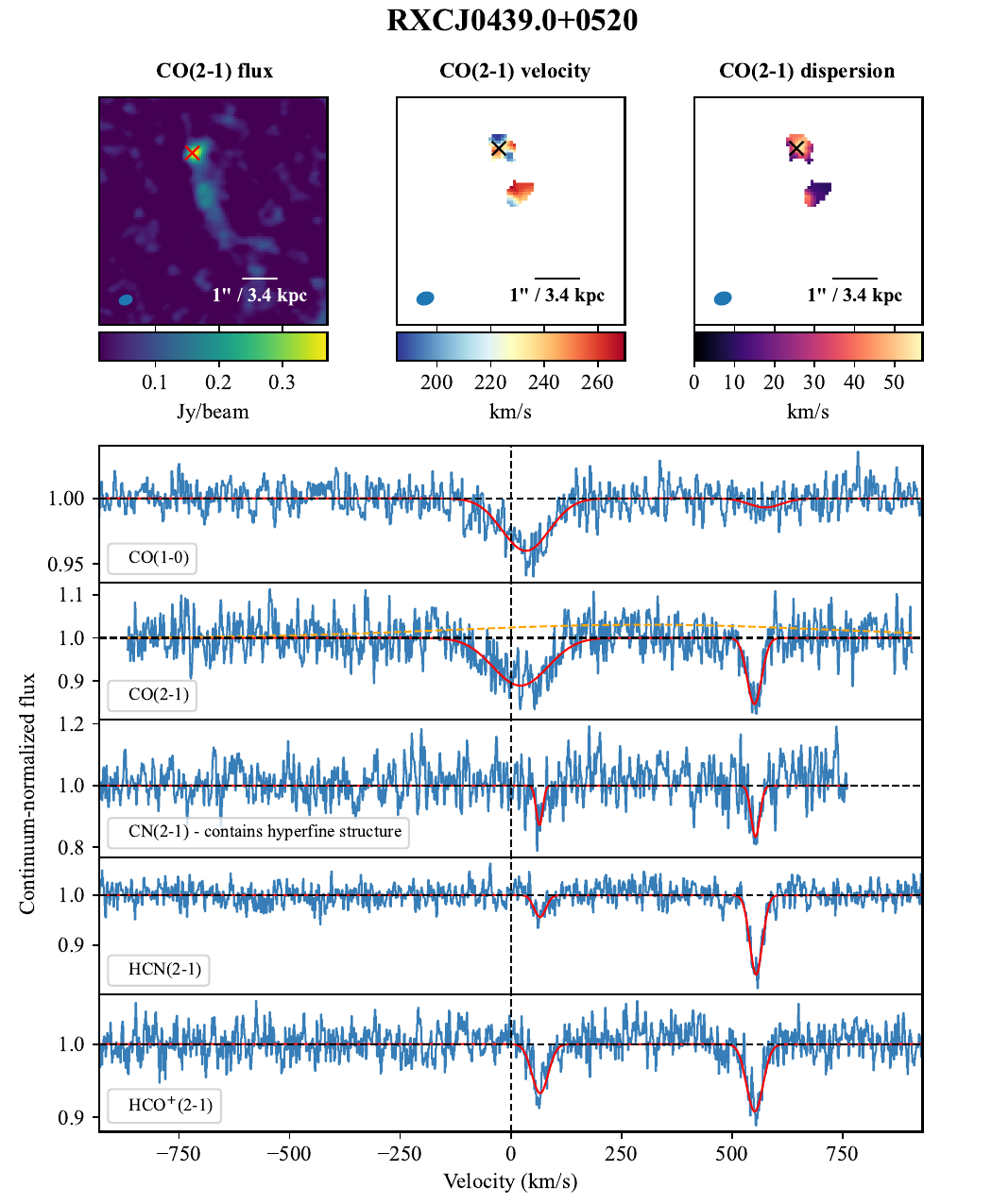}
    \caption{Intensity, velocity and velocity dispersion maps of RXCJ0439.0+0520 made from CO(2-1) emission. Crosses indicate the location of the radio continuum source, against which the spectra of several different molecular lines are extracted and shown below. The dashed yellow line shows subtracted CO emission.}
    \label{fig:J0439_maps_and_spectra}
\end{figure*}

\begin{figure*}
	\includegraphics[width=\textwidth]{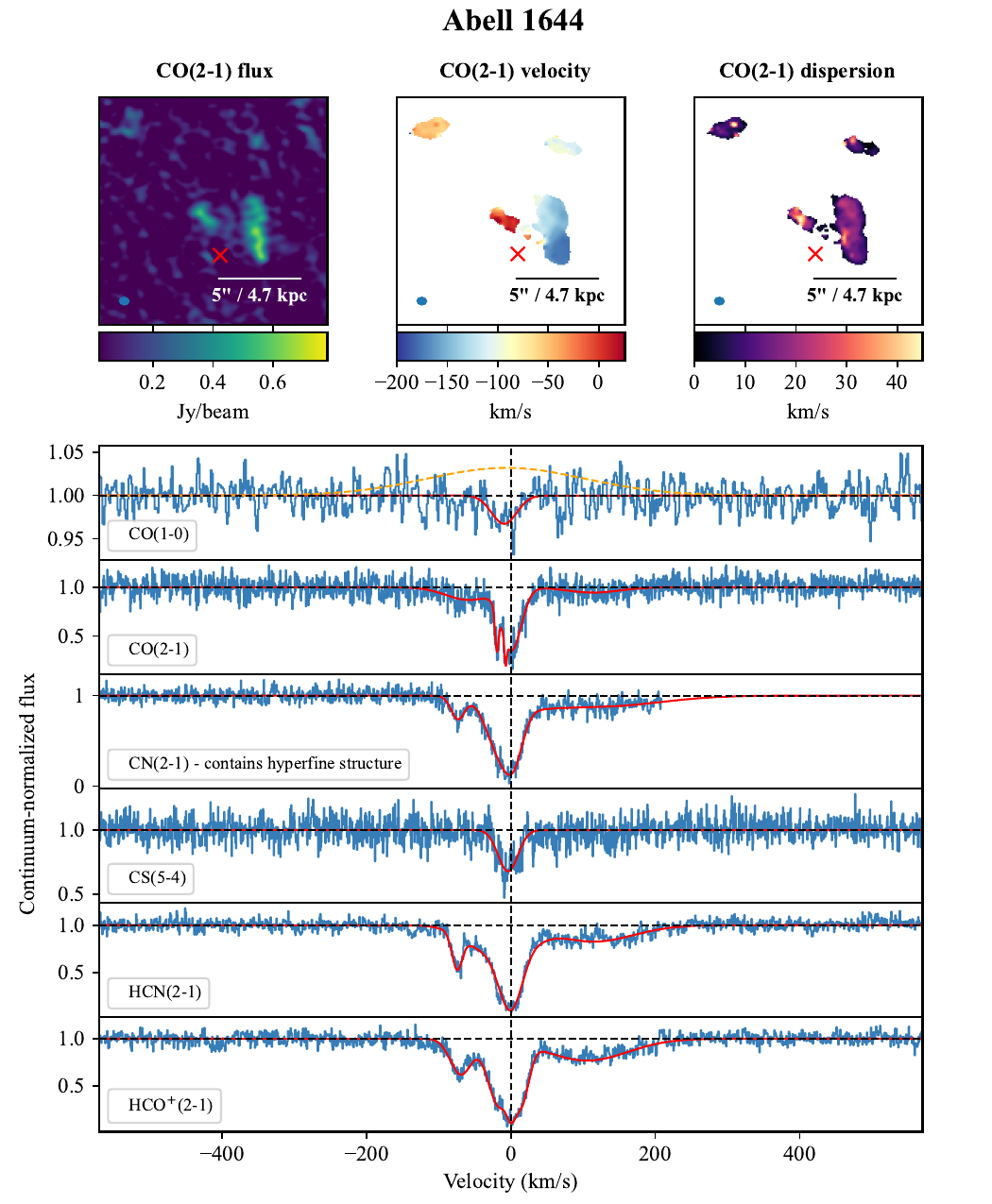}
    \caption{Intensity, velocity and velocity dispersion maps of Abell 1644 made from CO(2-1) emission. Crosses indicate the location of the radio continuum source, against which the spectra of several different molecular lines are extracted and shown below. The dashed yellow line shows subtracted CO emission. Note that the beam size of the CO(1-0) observation is much larger than in CO(2-1). The emission subtracted from the CO(1-0) spectrum is therefore from the more extended structures visible in the CO(2-1) map, and not from closer to the radio core.}
    \label{fig:A1644_maps_and_spectra}
\end{figure*}

\begin{figure*}
	\includegraphics[width=\textwidth]{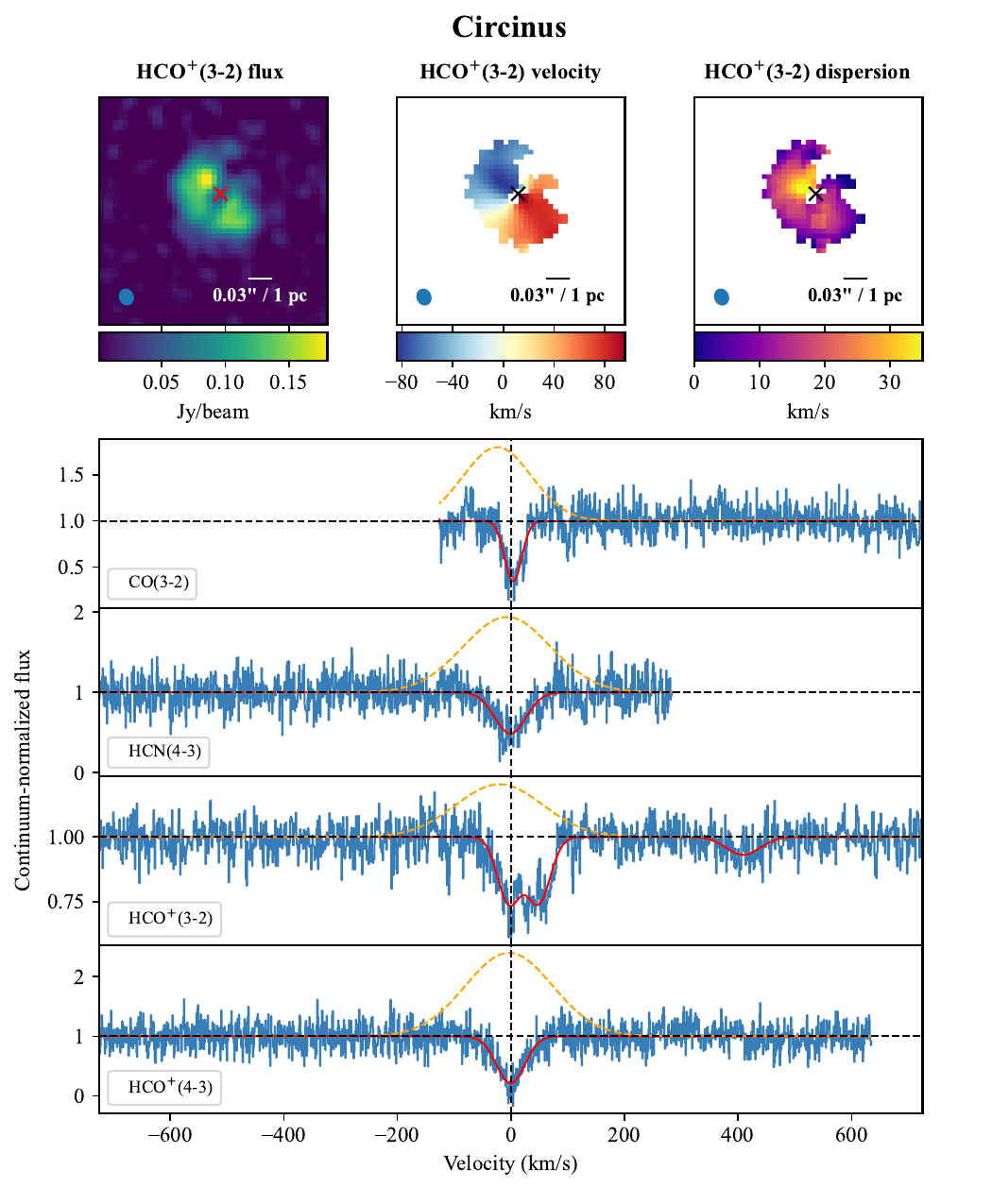}
    \caption{Intensity, velocity and velocity dispersion maps of Circinus made from HCO$^{+}$(3-2) emission. Crosses indicate the location of the radio continuum source, against which the spectra of several different molecular lines are extracted and shown below. Dashed yellow lines show the emission subtracted from each spectrum.}
    \label{fig:Circinus_maps_and_spectra}
\end{figure*}

\begin{figure*}
	\includegraphics[width=\textwidth]{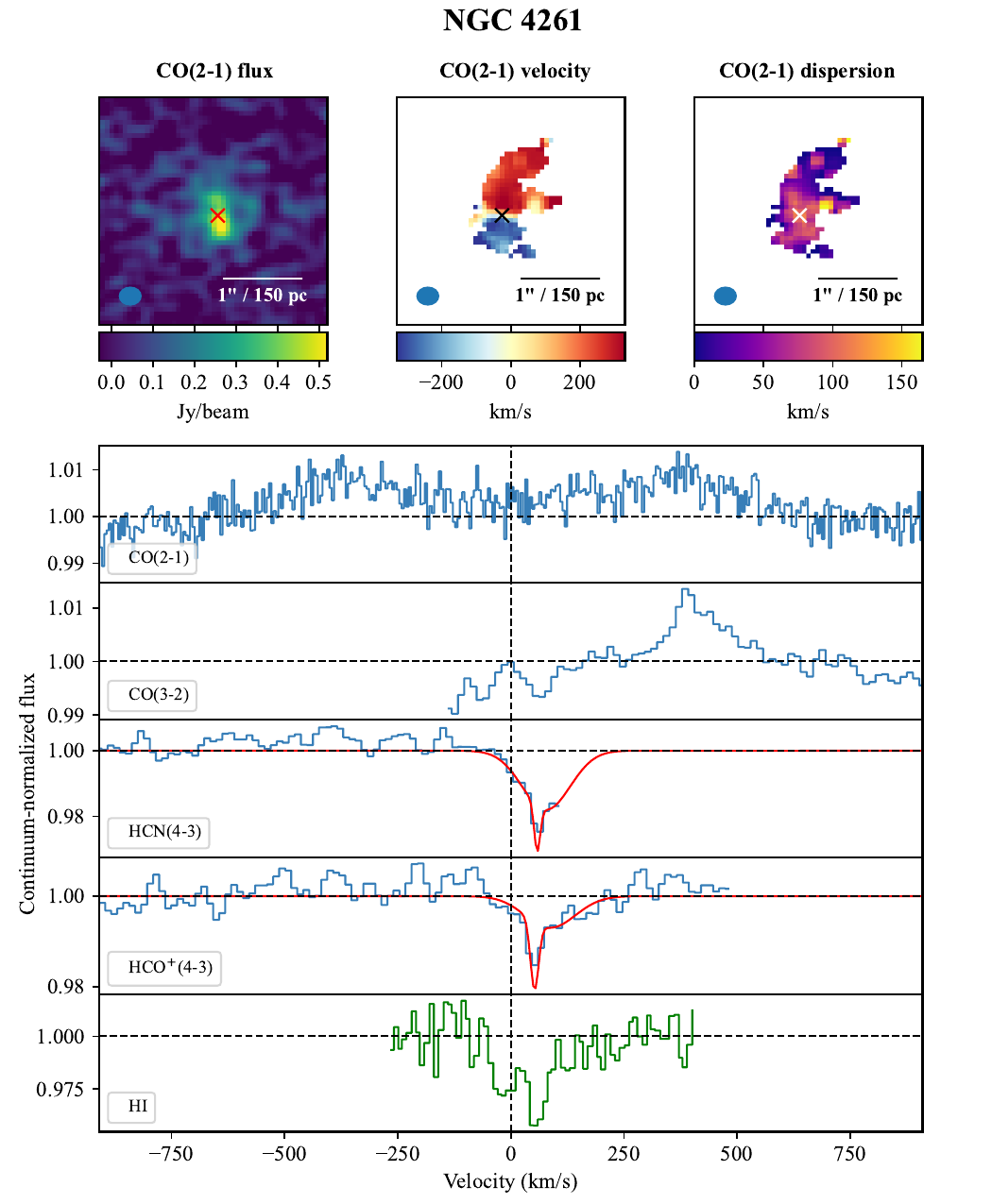}
    \caption{Intensity, velocity and velocity dispersion maps of NGC 4261 made from CO(2-1) emission. Crosses indicate the location of the radio continuum source, against which the spectra of different molecular lines are extracted and shown below. An H\small{I} spectrum also extracted against the radio core is shown in green \citep{Jaffe1994}.}
    \label{fig:NGC4261_maps_and_spectra}
\end{figure*}

\begin{figure*}
	\includegraphics[width=\textwidth]{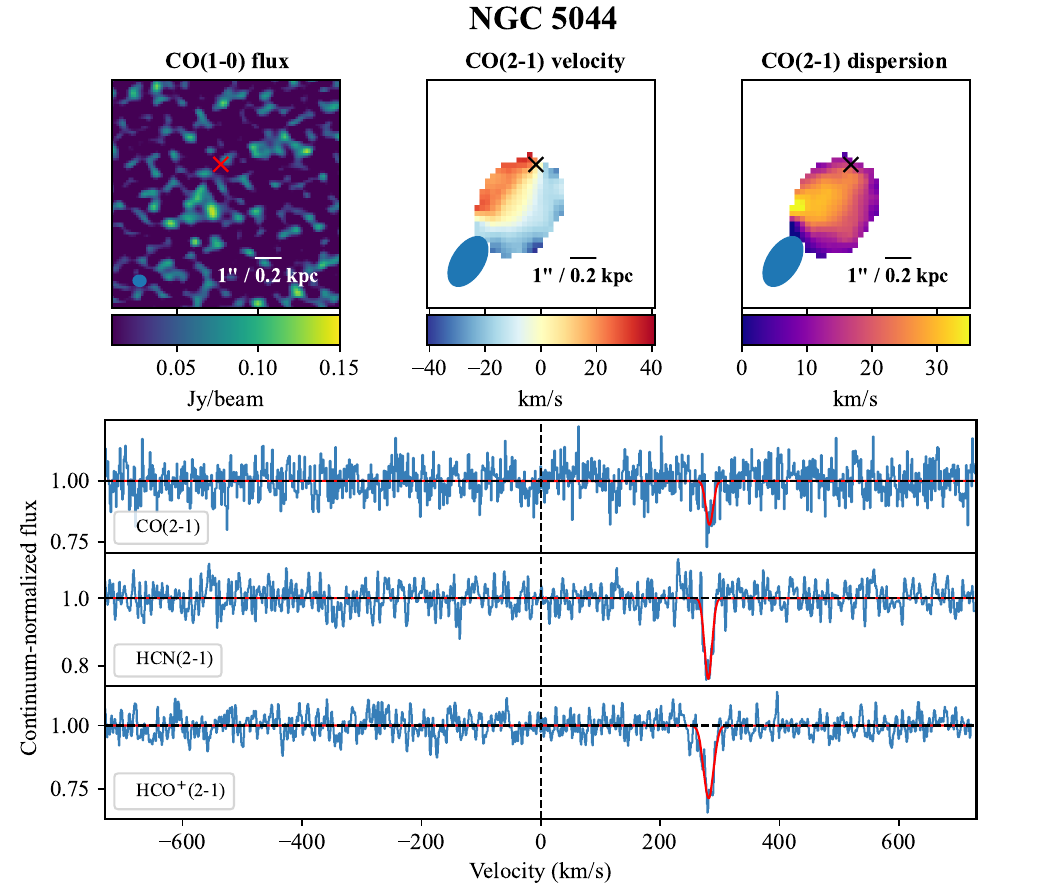}
    \caption{Intensity, velocity and velocity dispersion maps of NGC 5044 made from CO(1-0) and CO(2-1) emission. Crosses indicate the location of the radio continuum source, against which we extract spectra of three molecular lines.}
    \label{fig:NGC5044_maps_and_spectra}
\end{figure*}

\begin{figure*}
	\includegraphics[width=\textwidth]{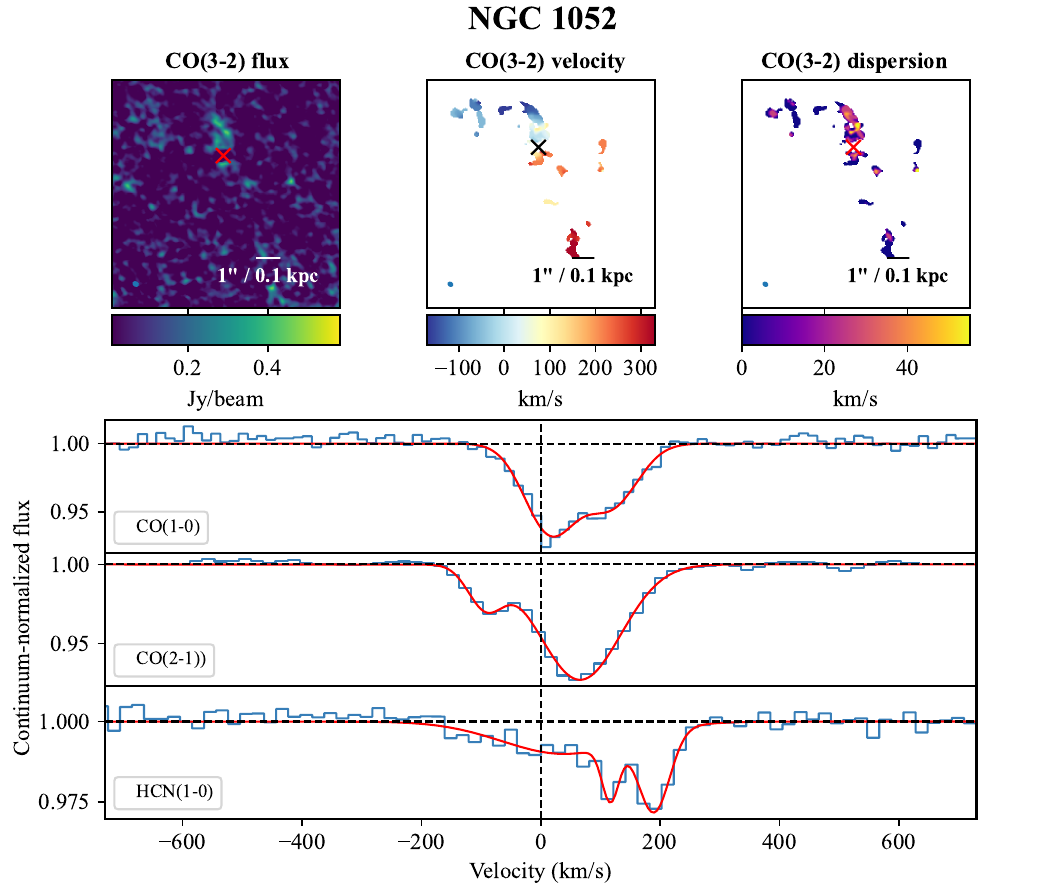}
    \caption{Intensity, velocity and velocity dispersion maps of NGC 1052 made from CO(3-2) emission. Crosses indicate the location of the radio continuum source, against which the spectra of three different molecular lines are extracted and shown below.}
    \label{fig:NGC1052_maps_and_spectra}
\end{figure*}

\begin{figure*}
	\includegraphics[width=\textwidth]{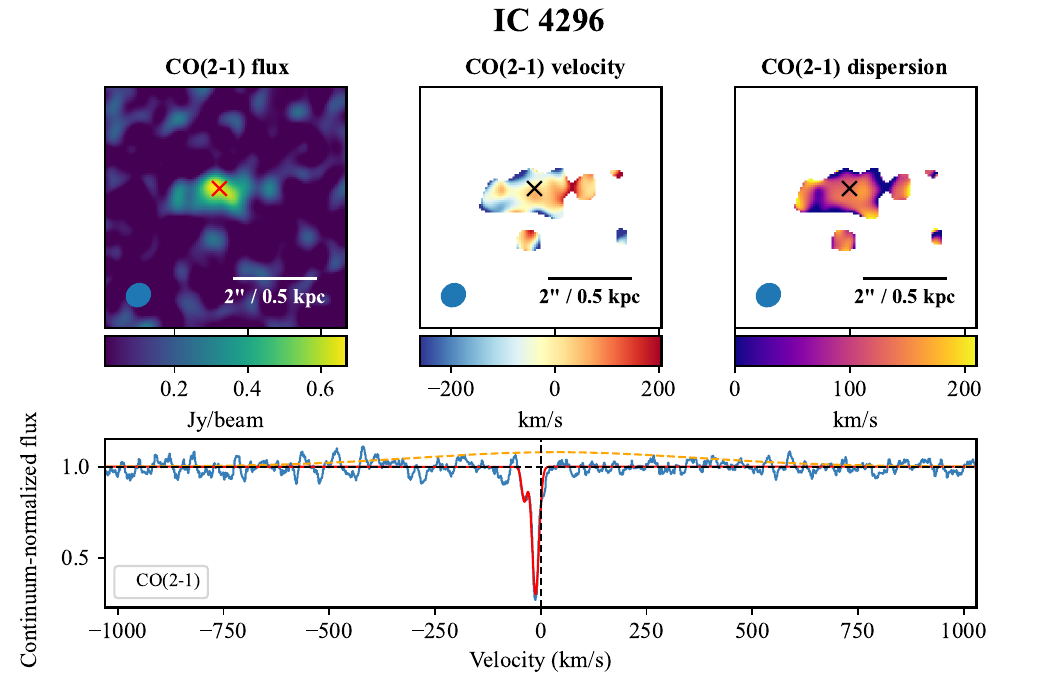}
    \caption{Intensity, velocity and velocity dispersion maps of IC 4296 made from CO(2-1) emission. Crosses indicate the location of the radio continuum source, against which the spectrum of CO(2-1) is extracted and shown below. The dashed yellow line shows subtracted CO emission.}
    \label{fig:IC4296_maps_and_spectra}
\end{figure*}

\begin{figure*}
	\includegraphics[width=\textwidth]{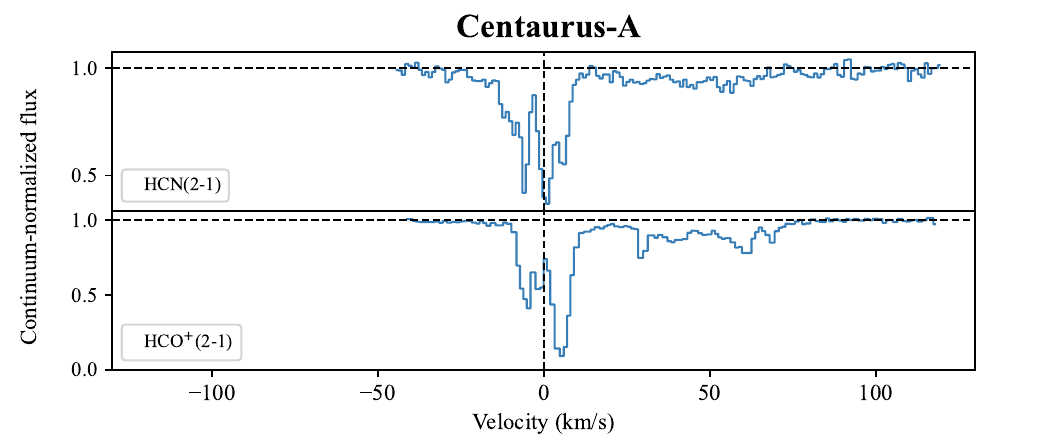}
    \caption{Spectra extracted from along the line of sight to the central radio continuum source of Centaurus-A, from \citep{Wiklind1997a}. Fits to the spectra can also be seen in that paper. Note that the velocity range of this spectrum is much narrower than in many of the preceding figures.}
    \label{fig:Centaurus-A_maps_and_spectra}
\end{figure*}

\begin{table}
\caption{Implied excitation temperatures, $T_{\textnormal{ex}}$. RXCJ0439.0+0520 has two clearly disconnected absorption features sufficiently strong in both spectra to calculate individual excitation temperatures (the first values given are for the 0\,km/s absorption feature). $^{1}$ For Hydra-A, we estimate the excitation temperature from the higher sensitivity HCO$^{+}$(2-1) and (1-0) observations \citep{Rose2020}.}
	\centering
	\begin{tabular}{lcc}	
\hline
& $\frac{\int\tau_{21}\,\textnormal{dv}}{\int\tau_{10}\,\textnormal{dv}}$ &  $T_{\textnormal{ex}}$ (K) \\
\hline
NGC 6868 & 7.1$_{-1.8}^{+1.8}$ & -15$_{-15}^{+4}$ \\ \\
S555 & 9.5$_{-0.8}^{+0.8}$ & -10.0$_{-1.1}^{+0.9}$ \\ \\
Hydra-A & 3.3$_{-0.3}^{+0.2}$ & $^{1}$11.1$_{-1.3}^{+1.4}$ \\ \\
Abell 2390 & 10.7$_{-1.4}^{+1.7}$ & -8.9$_{-1.4}^{+1.1}$ \\ \\
RXCJ0439.0+0520 & 3.4$_{-0.3}^{+0.3}$, 10.6$_{-1.2}^{+1.3}$ & 48$_{-16}^{+44}$, -9.2$_{-1.3}^{+0.9}$ \\ \\
Abell 1644 & 37$_{-9}^{+17}$ & -4.1$_{-0.5}^{+0.5}$ \\ \\
Abell 2597 & >5 & - \\ \\
NGC 5044 & >2 & - \\ \\
NGC 1052 & 1.2$_{-0.1}^{+0.1}$ & 5.9$_{-0.5}^{+0.5}$ \\ \\
\hline
	\end{tabular}
    \label{tab:excitation_temperatures}
\end{table}

\begin{table*}
	\centering
	\caption{Column densities estimated at 5\,K. The temperature used to calculate the column densities is the dominant uncertainty, so we do not give errors below.}
	\label{tab:gas_densities}
	\begin{tabular}{lccccccr} 
		\hline
		 & \multicolumn{7}{c}{Absorption Column Density / cm$^{-2}$} \\
        \hline
		 & CO & CO & CN & HCN & HCO$^{+}$ & HCN & HCO$^{+}$ \\
        Source & (1-0) & (2-1) & (2-1) & (2-1) & (2-1) & (4-3) & (4-3) \\
		\hline
		NGC 6868 & $1\times 10^{16}$ & $1\times 10^{17}$ & $1\times 10^{15}$ & $1\times 10^{14}$ & $1\times 10^{14}$ & -  & - \\
		Abell S555 (absorption at $\approx$180\,km/s) & - & $6\times 10^{15}$ & $3\times 10^{14}$ & $8\times 10^{13}$ & $8\times 10^{13}$ & -  & - \\
        Abell S555 (absorption at $\approx$270\,km/s) & $6\times 10^{15}$ & $9\times 10^{16}$ & $6\times 10^{14}$ & $1\times 10^{14}$ & $2\times 10^{14}$ & -  & - \\
        Hydra-A & $4\times 10^{15}$ & $2\times 10^{16}$ & $1\times 10^{14}$ & $3\times 10^{13}$ & $6\times 10^{13}$ & -  & - \\
        Abell 2597 & - & $7\times 10^{16}$ & - & $8\times 10^{14}$ & $1\times 10^{15}$ & -  & - \\
        Abell 2390 & $6\times 10^{16}$ & $1\times 10^{18}$ & - & $1\times 10^{15}$ & $2\times 10^{15}$ & -  & - \\
        RXCJ0439.0+0520 (absorption at $\approx$40\,km/s) & $1\times 10^{16}$ & $4\times 10^{16}$ & $6\times 10^{13}$ & $5\times 10^{12}$ & $1\times 10^{13}$ & -  & - \\
        RXCJ0439.0+0520 (absorption at $\approx$550\,km/s) & $1\times 10^{16}$ & $1\times 10^{16}$ & $1\times 10^{14}$ & $2\times 10^{13}$ & $1\times 10^{13}$ & -  & - \\
        Abell 1664 & - & $1\times 10^{17}$ & $3\times 10^{15}$ & $6\times 10^{14}$ & $5\times 10^{14}$ & -  & - \\
        Circinus$^{*}$ & - & - & - & - & - & $2\times 10^{14}$ & $5\times 10^{14}$ \\
        NGC 4261 & - & - & -  & - & - & $1\times 10^{13}$ & $4\times 10^{12}$ \\
        NGC 5044 & - & $8\times 10^{15}$ & - & $1\times 10^{13}$ & $2\times 10^{13}$ & -  & - \\
        NGC 1052 & $2\times 10^{16}$ & $3\times 10^{16}$ & - & -  & - & -  & - \\
        IC 4296 & - & $1\times 10^{17}$ & - & - & - & -  & - \\
		\hline
        \multicolumn{3}{c}{$^{*}$Circinus also has a CO(3-2) column density of $1\times 10^{17}$ / cm$^{-2}$.}
	\end{tabular}
\end{table*}

\section{Molecular gas properties}

\subsection{Excitation temperatures}
\label{sec:excitation_temperatures}

The excitation temperature, $T_{\mathrm{ex}}$, of a two level system is defined in terms of the number of particles in the upper state, $n_{\mathrm{u}}$ and lower state, $n_l$:
\begin{equation}
\frac{n_{\mathrm{u}}}{n_{\mathrm{l}}}=\frac{g_{\mathrm{u}}}{g_{\mathrm{l}}} \exp \left(-\frac{h\nu}{k T_{\mathrm{ex}}}\right),
\end{equation}
where $g$ is the statistical weight of each state. The excitation temperature should not be confused with the more familiar thermal temperature. Rather, it is the temperature that would be expected to produce the ratio of particles in the different energy levels when local thermodynamic equilibrium applies.

For optically thin gas in local thermodynamic equilibrium, the \mbox{CO(1-0)} and \mbox{CO(2-1)} velocity integrated optical depths are related by
\begin{equation}
\label{eq:opacityratio}
\frac{\int \tau_{21} dv}{\int \tau_{10} dv} = 2 \frac{1 - e^{- h\nu_{21}/k T_{\textnormal{ex}}}}{e^{h\nu_{10}/k T_{\textnormal{ex}}} -1}\enspace ,
\end{equation}
where $h$ and $k$ are the Planck and Boltzmann constants, $\nu_{10}$ and $\nu_{21}$ are the rest frequencies of the \mbox{CO(1-0)} and \mbox{CO(2-1)} lines and $T_{\textnormal{ex}}$ is the excitation temperature \citep{Bolatto2003,Godard2010,Magnum2015}. According to Equation \ref{eq:opacityratio}, the ratio of CO(2-1)/CO(1-0) velocity integrated optical depths can be in the range of 0 to 4. These respectively correspond to the extreme cases where $k$T$_{\textnormal{ex}}<<\textnormal{h}\nu$ and $k$T$_{\textnormal{ex}}>>\textnormal{h}\nu$.

For each galaxy where we have CO(1-0) and CO(2-1) absorption, we estimate the ratios of the velocity integrated optical depths and the implied excitation temperatures. These are shown in Table \ref{tab:excitation_temperatures}. We generally estimate a single excitation temperature for each system's absorption, even though in many cases separate components are clearly present. This is because these components overlap significantly and are heavily degenerate. RXCJ0439.0+0520 is the exception, where the two components of absorption are separated by 500\,km/s. We therefore calculate two excitation temperatures. 

Many of the CO(2-1)/CO(1-0) ratios in Table \ref{tab:excitation_temperatures} fall outside the feasible range of 0 -- 4 and therefore give a negative excitation temperature. The excitation temperature may be negative when $n_{\mathrm{u}}/g_{\mathrm{u}} > n_{\mathrm{l}}/g_{\mathrm{l}}$, which is possible if there is pumping by a maser. However, this effect is not possible in CO. Therefore, one or more of the assumptions made in our calculations is incorrect.

One possibility is that the gas is not in local thermodynamic equilibrium, i.e. the excitation temperature may vary between the different rotational levels. If the CO(1-0) excitation temperature is 20\,K and the CO(2-1) excitation temperature is 5\,K, then a ratio in the velocity integrated optical depths of 5.6 is expected. This temperature difference is at the more extreme end of what could be expected, but is still far from explaining the ratios of over 10 which we see in the absorption lines of some systems.

A second assumption implicit in our calculations is that the size of the continuum source does not change between the frequencies of the CO(1-0) and CO(2-1) lines. This can alter the line of sight at different frequencies, changing how the clouds intercept the continuum. We explore this possibility in more detail in our discussion. In the mean time however, we proceed to estimate molecular column densities by assuming the ground state excitation temperature of the molecular gas is 5\,K for all of the molecular lines.

\subsection{Absorption column density estimates}

The line of sight column density, $N_{\textnormal{tot}}$, of an optically thin molecular absorption region is:

\begin{equation}
\label{eq:thin_colum_density}
N_{\textnormal{tot}}^{\textnormal{thin}} = Q(T) \frac{8 \pi \nu_{ul}^{3}}{c^{3}}\frac{g_{l}}{g_{u}}\frac{1}{A_{ul}} \frac{1}{ 1 - e^{-h\nu_{ul}/k T_{\textnormal{ex}}}}\int \tau_{ul}~dv~,
\end{equation}

where $Q$($T$) is the partition function, $c$ is the speed of light, $A_{ul}$ is the Einstein coefficient of the observed transition and $g$ the level degeneracy, with the subscripts $u$ and $l$ representing the upper and lower levels \citep{Godard2010,Magnum2015}. 

The assumption of optically thin absorption in Equation \ref{eq:thin_colum_density} is inappropriate for some of our data. To account for this, a correction factor from \citet{Magnum2015} can be applied to give the following more accurate column density:

\begin{equation}
N_{\textnormal{tot}} = N_{\textnormal{tot}}^{\textnormal{thin}} \frac{\tau}{1-\exp({-\tau})}.
\label{eq:true_column_densities}
\end{equation}
This gives the column density of the absorbing molecule as a number density per square cm. 

Table \ref{tab:gas_densities} shows the column densities estimated using each object's molecular absorption lines. By far the most dominant error in our calculations of the column densities is the gas excitation temperature, $T_{\textnormal{ex}}$. We therefore use only the central estimate of the velocity integrated optical depth in our calculations.

\subsection{Molecular mass estimates}
\label{sec:masses}
The molecular mass associated with CO emission can be estimated using the following relation from \citet{Bolatto2013}:

\begin{equation}
\begin{split}
\textnormal{M}_{\text{mol}} = \frac{1.05\times 10^{4}}{F_{ul}} \left( \frac{X_{\text{CO}}}{2\times 10^{20}\frac{\text{cm}^{-2}}{\text{K km s}^{-1}}}\right)\left( \frac{1}{1+z}\right) \\ \times \left( \frac{S_{\text{CO}} \Delta v}{\text{Jy km s}^{-1}}\right) \left( \frac{D_{\text{L}}}{\text{Mpc}}\right)^{2}\, \textnormal{M}_{\odot},
\end{split}
\label{eq:massequation}
\end{equation}
where $M_{\text{mol}}$ is the mass of molecular hydrogen, $X_{\text{CO}}$ is a CO-to-H$_{2}$ conversion factor, $z$ is the source's redshift, $S_{\text{CO}} \Delta v$ is the emission integral and $D_{\text{L}}$ is the luminosity distance in Mpc. $F_{ul}$ is an approximate conversion factor for the expected flux density ratios of the CO(1-0) and CO(2-1) lines, where $u$ and $l$ represent the upper and lower levels. For CO(1-0), $F_{10}=1$ and for CO(2-1), $F_{21}=3.2$.

To ensure our mass estimates are comparable with similar studies, we use the standard Milky Way CO-to-H$_2$ conversion factor of \mbox{$X_\text{CO} = 2 \times 10^{20}$ cm$^{-2}$ (K km s$^{-1}$)$^{-1}$} in our calculations. However, this is a considerable source of uncertainty. In the brightest cluster galaxy RXJ0821+0752 for example, \citet{Vantyghem2017} use $^{13}$CO emission to constrain the conversion factor and find it to be roughly half of the Galactic value.

The molecular mass we find for each source is given in Table \ref{tab:molecular_properties}. In NGC 4261, we find a mass of $2.1\pm 0.3 \times 10^{7}$M$_{\odot}$. This is higher than the $1.1 \times 10^{7}$M$_{\odot}$ found by \citet{Boizelle2021}. Using the `product' data (a preliminary image delivered with all ALMA data) we also find this value. However, this image appears to have had its continuum over-subtracted, so we have calculated the mass after performing our own continuum subtraction.

\section{Discussion}

\subsection{Velocity distribution of molecular absorption lines -- evidence of cold gas inflow}

\begin{figure*}
	\includegraphics[width=\textwidth]{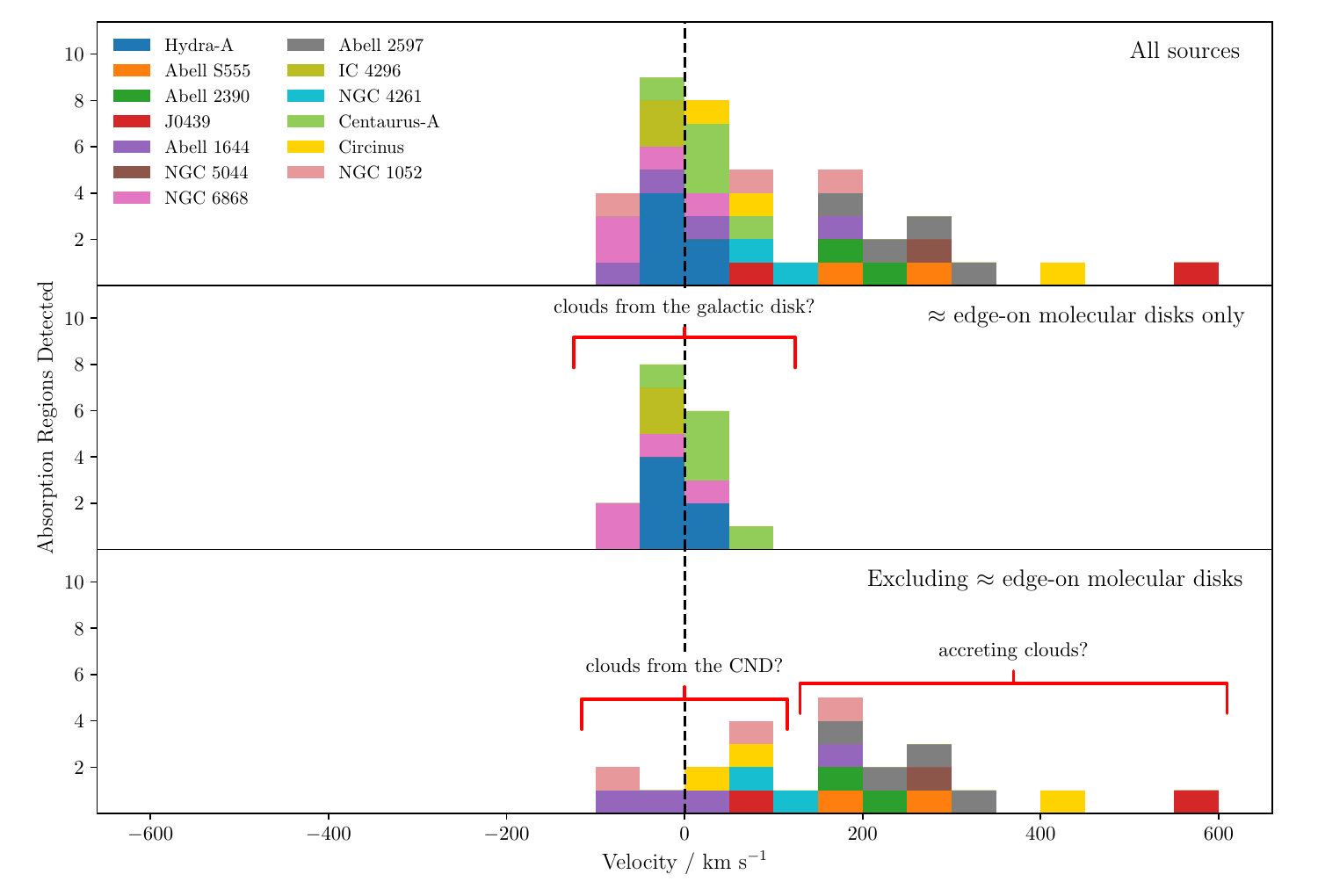}
    \caption{\textbf{Upper}: The velocities of all intrinsic molecular absorption regions detected (known to the authors) with complementary high resolution molecular emission observations. The bias for positive velocities implies movement of the absorbing molecular gas towards the galaxy centres. Each galaxy's systemic velocity has an error of 30\,km/s or less. To ensure a fair comparison, we show the number of individually resolved absorption regions after all spectra are smoothed into 5\,km/s spectral bins. \textbf{Middle}: Same as above, but only for galaxies with a disk of molecular gas inclined at $45^{\circ}<i\leq90^{\circ}$ (i.e. edge-on or close to edge-on). All absorption regions here are narrow ($\sigma$<9km/s), except one in each of NGC 6868 and Centaurus-A. \textbf{Lower:} Same as upper, but excluding galaxies with a disk of molecular gas inclined at $45^{\circ}<i\leq90^{\circ}$.}
    \label{fig:Histogram}
\end{figure*}

\begin{figure}
	\includegraphics[width=0.98\columnwidth]{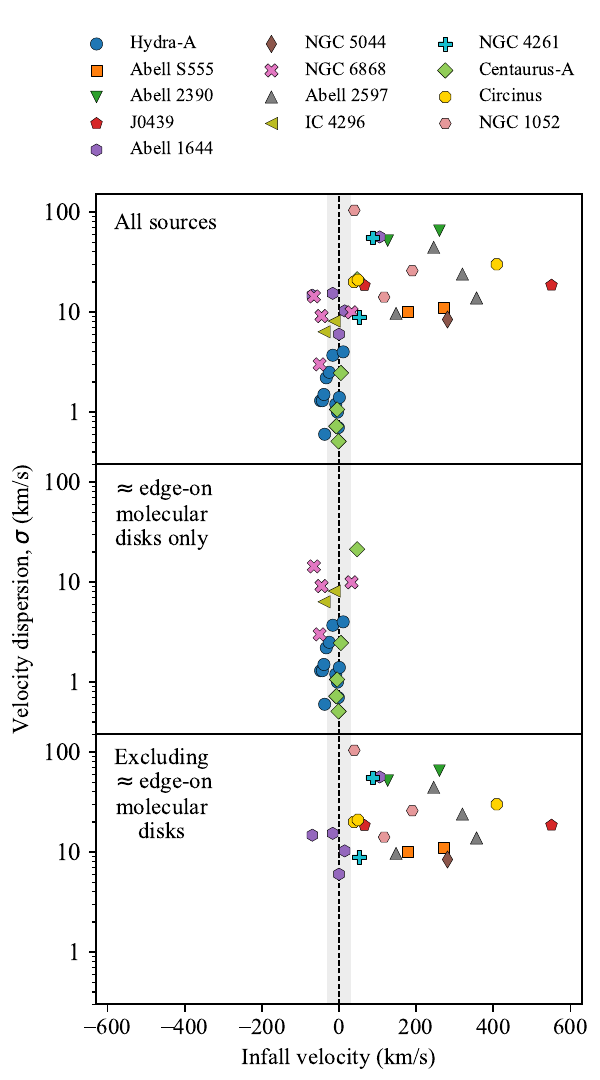}
    \caption{The infall velocity of molecular clouds versus their velocity dispersion, using the molecular line in which each system's absorption is best detected. In the top plot, we show the absorption regions detected in all sources. In the middle, we show those detected in molecular disks inclined at $45^{\circ}<i\leq90^{\circ}$ (i.e. edge-on or close to edge-on). This orientation makes our effective line of sight through the disk as high as possible, maximising the likelihood of absorption from its molecular clouds. At the bottom, we show absorption regions from galaxies which do not have a molecular disk inclined at $45^{\circ}<i\leq90^{\circ}$. In these cases, there is a strong bias for redshifted absorption with high velocity dispersions. The grey band indicates the typical uncertainty in the recession velocity of each source.}
    \label{fig:infall_vs_sigma}
\end{figure}

\begin{figure}
	\includegraphics[width=\columnwidth]{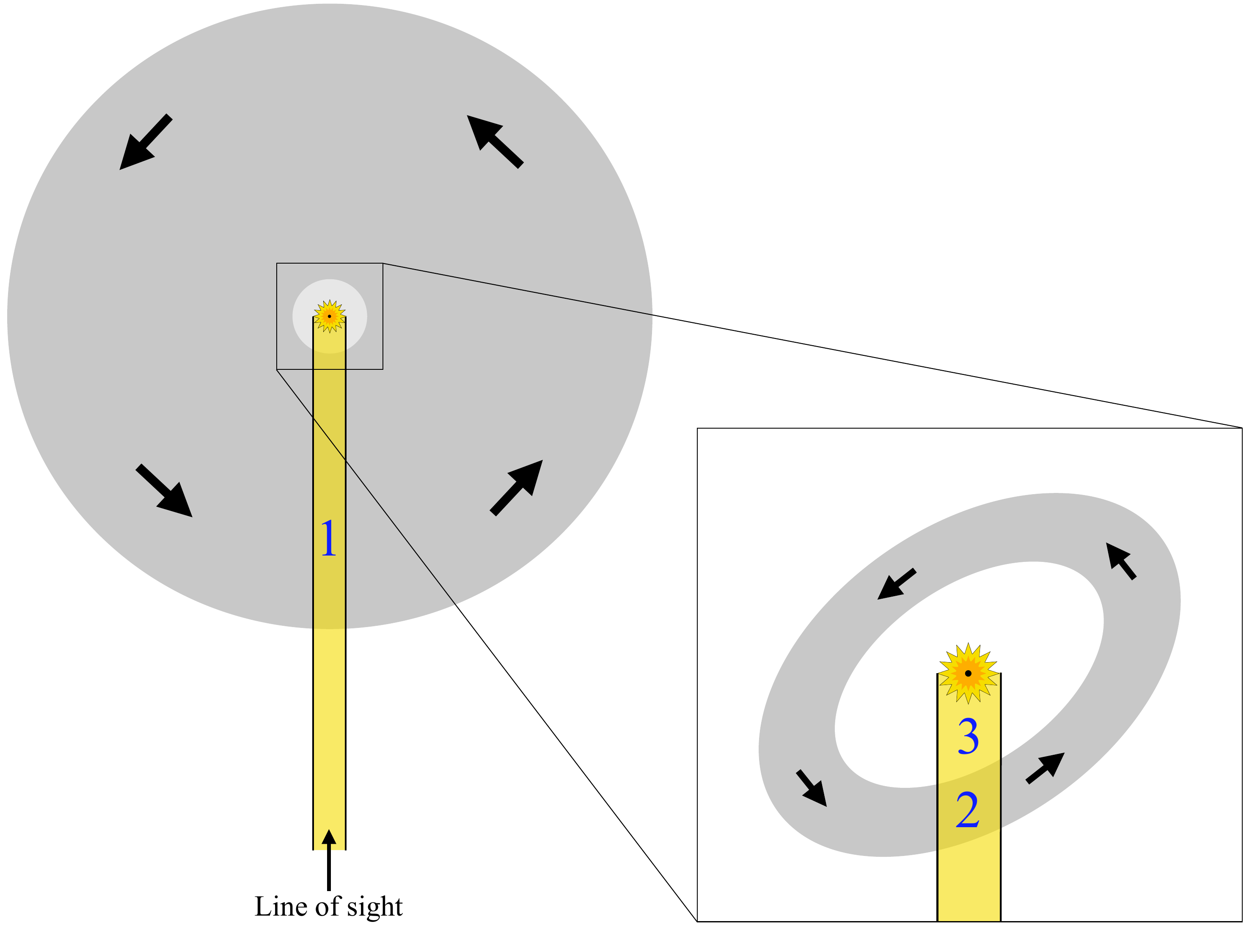}
    \caption{A bird's eye view of our line of sight to the radio continuum of an edge-on disk galaxy. Gas clouds producing molecular absorption lines may be classified as coming from three regions, as labelled above. Clouds from region 1 lie within the galaxy's large scale distribution of molecular gas, which is readily visible via molecular emission. Region 2 clouds are within the circumnuclear disk surrounding the supermassive black hole. The elliptical nature of these orbits can induce small red and blueshifted line of sight velocities. Clouds from region 3 originate from the circumnuclear disk and are likely falling towards the galaxy centre. This may be due to cloud-cloud collisions which have caused them to lose angular momentum. If the line of sight changes such that the galactic disk is no longer seen edge-on, the relative probability of absorption coming from regions 2 and 3 increases.}
    \label{fig:schematic_of_absorption_regions}
\end{figure}

\begin{figure*}
	\includegraphics[width=0.8\textwidth]{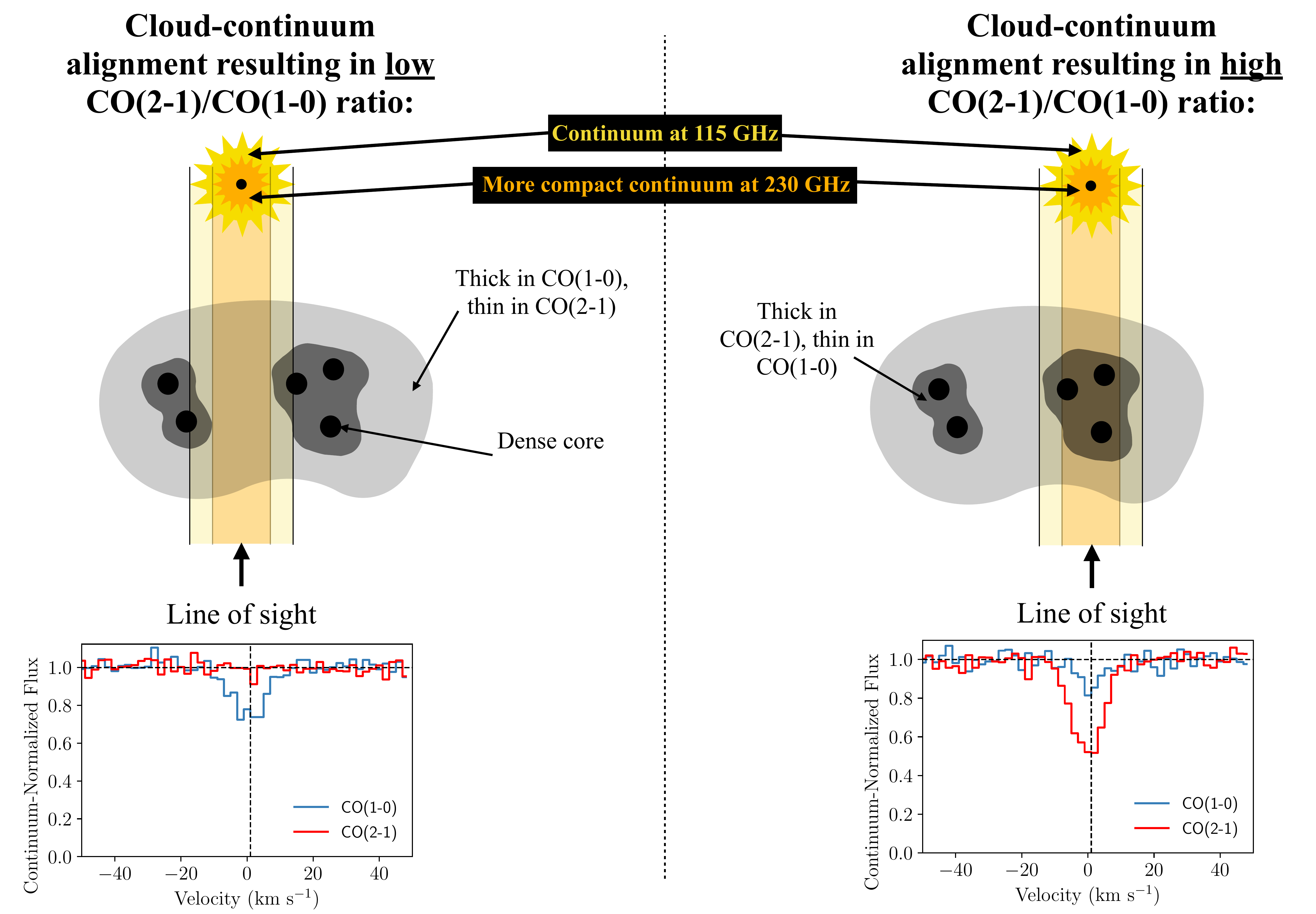}
    \caption{For a background continuum whose size varies with frequency, the relative strengths of optically thin absorption from different molecular lines may be influenced by how the cloud is positioned relative to the line of sight. Here we show the line of sight to the 115\,GHz continuum in yellow, and the more compact 230\,GHz continuum in orange. For the alignment on the left, our line of sight passes through no CO(2-1) thick gas, giving a low CO(2-1)/CO(1-0) ratio. For the alignment on the right, the line of sight to the 230\,GHz continuum passes through a considerable column density of CO(2-1) thick gas, giving a high CO(2-1)/CO(1-0) ratio.}
    \label{fig:cloud_diagram}
\end{figure*} 

In Figure \ref{fig:Histogram}, we show histograms of the absorption line velocities for the galaxies shown in Figures \ref{fig:NGC6868_maps_and_spectra} -- \ref{fig:Centaurus-A_maps_and_spectra}. The top panel shows all intrinsically absorbing systems known to us with complementary high resolution molecular emission line observations. In the middle panel, we show the absorption regions found in galaxies possessing a molecular disk with an inclination angle of $45^{\circ}<i\leq90^{\circ}$ (where $i=90^{\circ}$ equates to an edge-on disk). In the lower panel, we show those which have no such disk. When making these histograms, we use the molecular species in which each source's absorption lines are best resolved. This is typically HCO$^{+}$, but is sometimes another species\footnote{For each galaxy, we use: NGC6868 -- HCO$^{+}$(2-1), Abell S555 -- HCO$^{+}$(2-1), Hydra-A -- HCO$^{+}$(2-1), Abell 2597 -- HCO$^{+}$(2-1), Abell 2390 -- HCO$^{+}$(2-1), RXCJ0439.0+0520 -- HCO$^{+}$(2-1), Abell 1644 -- HCO$^{+}$(2-1), Circinus -- HCO$^{+}$(3-2), NGC 4261 -- HCO$^{+}$(4-3), NGC 5044 -- HCO$^{+}$(2-1), NGC 1052 -- HCN(1-0), IC 4296 -- CO(2-1), Centaurus-A -- HCN(2-1)}. We also limit the velocity resolution to 5\,km/s to ensure the spectra of different sources are comparable.

In the top panel, around two-thirds of the absorption regions detected are evenly distributed within 100\,km/s of the zero velocity point. The remaining third are spread across a redshifted wing between 100 and 550 km/s. The distribution of velocities is asymmetric, with the equally blueshifted side lacking any absorption regions. For edge-on or close to edge-on molecular disks, the absorption regions are spread symmetrically about 0\,km/s. When excluding sources with an edge-on or close to edge-on molecular disk, the bias for redshifted absorption is particularly strong.

Due to the way absorption lines are formed, each gas cloud's direction of motion with respect to the nucleus is known without ambiguity. Since an absorbing cloud must lie between the observer and the continuum, redshift (i.e. a positive velocity) indicates movement towards the galaxy centre. The only exception to this is when the relative velocity is near zero and comparable to the velocity error, typically 10-30\,km/s.

The high degree of asymmetry in Figure \ref{fig:Histogram} therefore provides strong evidence of molecular gas migrating towards galaxy centres in significant quantities, particularly when excluding edge-on and close to edge-on disks. In Figure \ref{fig:infall_vs_sigma}, we also show the infall velocity of the absorption regions versus their velocity dispersion. This shows two distinct populations of molecular clouds. In disks, we see narrow absorption features with no net movement towards the galaxy centre. Where our line of sight does not pass through a disk, the absorption is caused by high velocity dispersion clouds. Most of these are moving towards the galactic centre along the line of sight at hundreds of km/s.

\subsection{The location of absorbing molecular clouds}
\label{sec:location_of_clouds}
With the velocity distribution shown in Figure \ref{fig:Histogram}, it is interesting to consider the different regions and environments within a galaxy where absorbing molecular clouds may lie, and how this would affect their properties. This is particularly important given that we lack information regarding their proximity to their galaxy's centre, even if we know their RA, Dec and line of sight velocity with certainty.

In Figure \ref{fig:schematic_of_absorption_regions} we show a simplified bird's eye view of our line of sight to a galaxy with an edge-on disk of molecular gas. Perhaps the most obvious region in which clouds may lie is within the large, typically kpc scale distribution of molecular gas (which will not necessarily be a disk). We label this as region 1.

These large scale gas distributions are normally studied in ensemble via emission. However, in the Milky Way it is possible to study molecular clouds via emission on smaller mass scales due to their proximity. In the Galaxy, cold gas tends to be clumpy and assembled into clouds of 10s to 1000s of solar masses \citep{Gong2016, RomanDuval2010}. There is also no known net flow of gas towards or away from the galaxy centre, and any orbital velocities will be perpendicular to our line of sight. These properties, combined with their relatively stable environment, give the clouds low line of sight velocities and velocity dispersions of a few km/s or less. Therefore, if the clouds we detect via absorption are from similar kpc scale distributions of molecular gas, they are likely to produce narrow absorption features with low line of sight velocities.

For a large scale distribution of gas, the probability of its constituent clouds aligning with the continuum is heavily dependent on our line of sight. The probability will be maximised for an edge on disk e.g. Hydra-A. In this case, any change in inclination will reduce the line of sight column density of molecular gas and thus reduce the chance of a cloud-continuum alignment. Therefore, when we show the absorption regions detected in edge-on or close to edge-on disks in the central panels of Figures \ref{fig:Histogram} and \ref{fig:infall_vs_sigma}, the probability that the absorption has arisen from gas within each galaxy's disk is maximised.

Molecular absorption may also originate from the circumnuclear disk, labelled `2' in Figure \ref{fig:schematic_of_absorption_regions}. Although this region is of much lower mass than the galaxy-wide distribution of molecular gas, its proximity to the continuum source means an individual cloud's chance of aligning with the continuum is higher. 

The circumnuclear disk is a highly energetic region which surrounds an AGN, and is composed of molecular and atomic gas. In the Milky Way for example, the circumnuclear disk has an inner radius of 1.5\,pc and extends out to around 7\,pc, comparable to the supermassive black hole's sphere of influence \citep{Oka2011}. Due to the highly energetic nature of this region, its molecular clouds would likely have a high velocity dispersion, which would be imprinted onto their molecular absorption lines. 

A galaxy's circumnuclear disk is also typically composed of gas on elliptical orbits \citep{Solanki2023, Izumi2023}. If these are aligned such that neither the semi-major or semi-minor axes are parallel to the line of sight, an apparent line of sight velocity can arise. In this way, elliptical orbits may induce a line of sight velocity of up to 200\,km/s \citep{Rose2024}. However, the orientation of the orbit will rarely maximise the apparent line of sight velocity. Additionally, an inclination of the elliptical orbit will reduce it further by a factor of $\sin{i}$. Therefore, the red and blueshifted velocities induced by this effect will typically be 100\,km/s or less. Importantly, the apparent line of sight velocities from this effect have an equal probability of being red or blueshifted, since the sign of the velocity simply depends on observed direction of rotation. Therefore, the symmetric cluster of absorption regions in the lower panels of Figure \ref{fig:Histogram} and \ref{fig:infall_vs_sigma}, which have a roughly symmetric distribution of velocities between -100 and 100\,km/s, are likely to be from this region.

Numerical simulations suggest that the vast reservoirs of molecular gas in massive galaxies exist, at least in part, as a population of relatively small clouds within radii of a few tens to hundred of parsecs \citep{Pizzolato2005, Gaspari2015, Gaspari2018}. These clouds are expected to undergo inelastic collisions, leading to angular momentum loss and their funnelling towards the central supermassive black hole, eventually providing it with fuel \citep{Vollmer2004}. For regions with a high velocity dispersion like the circumnuclear disk, these cloud-cloud collisions are particularly likely. Therefore, regions 2 and 3 of Figure \ref{fig:schematic_of_absorption_regions} may be composed of clouds originating from the circumnuclear disk, which have undergone collisions and are migrating towards the supermassive black hole. As such, absorption from this region may be expected to be have a wide velocity dispersion, as well as being redshifted. To estimate the possible speeds these clouds may attain, we can consider them to be in freefall. If all the lost gravitational potential energy is converted to kinetic energy, then:

\begin{equation}
\label{eq:infall_velocity}
    v = \left[ \textnormal{2G}M \left( \frac{1}{r_{1}} - \frac{1}{r_{0}} \right) \right]^{0.5},
\end{equation}

where $v$ is the cloud velocity after falling from stationary at a height $r_{0}$ to $r_{1}$ towards a mass $M$. 

The exact radii at which clouds will begin to fall is uncertain. However, we can make estimates based on the size of the circumnuclear disk detected in Circinus. At the centre of the galaxy, \citet{Izumi2023} infer the presence of a $1.7\times10^{6}\,\textnormal{M}_{\odot}$ supermassive black hole (plus a molecular mass one order of magnitude lower). If freefalling from 2\,pc to 1\,pc, a velocity of 90\,km/s is reached. At 0.5\,pc, the velocity reaches 150\,km/s. A velocity matching that of the 420\,km/s HCO$^{+}$(3-2) absorption feature would be reached at a radius of 0.09\,pc ($1.9\times 10^{4}$\,AU). 

The clusters in our sample with the more massive central galaxies e.g. Abell 2390 and Abell 2597 have black hole masses around three orders of magnitude higher than Circinus. Depending on the size of the circumnuclear disks, infall may therefore be considerably faster.

\subsection{The fate of infalling molecular clouds}

Figures \ref{fig:Histogram} and \ref{fig:infall_vs_sigma} represent convincing evidence of cold molecular gas flowing into galactic centres at hundreds of km/s. However, the infall of these clouds may not be modelled well by freefall. Instead, with some initial angular momentum, they are likely to enter stable elliptical orbits rather than falling directly onto the supermassive black hole and providing the active galactic nucleus with fuel. The radii of these stable orbits is likely extremely low, and probably sub-parsec. At this proximity to the galaxy centre, the molecular gas is likely to rapidly change into an atomic and/or ionised phase. The wide velocity dispersion of the HI gas in Abell 2597 and Abell 2390 (Figures \ref{fig:A2597_maps_and_spectra} and \ref{fig:A2390_maps_and_spectra}) may be a manifestation of this effect, with the outer layers of the infalling clouds being evaporated and their velocity dispersion increasing. Ultimately, a fraction of the gas we see migrating towards the galactic centres may contribute to the fuelling of the active galactic nucleus, with the rest likely to be ejected by radio jets and lobes in what are often seen as ionised outflows \citep[e.g.][]{Harrison2018, Choi2020, Laha2021, Singha2021}.

\subsection{Cloud structure}

The breakdown in our excitation temperature calculations of section \ref{sec:excitation_temperatures} and the resulting negative temperatures imply a false premise in our assumptions of thermodynamic equilibrium, homogeneous clouds and a continuum source which does not change in size with frequency. However, we ruled out a lack of non-local thermodynamic equilibrium as the cause because its effect is too small. Instead, a continuum that changes in size with frequency, combined with the hierarchical structure of the molecular clouds, could explain the otherwise impossible CO(2-1)/CO(1-0) molecular line ratios.

Molecular gas concentrations in massive galaxies are predicted to have a hierarchical structure. This consists of dense, star forming cores up to 0.1\,pc in diameter, $\sim 0.1$\,pc wide filaments, $\sim 1$\,pc wide clumps and $\sim 1-10$\,pc sized clouds \citep{Bergin2007}. In Figure \ref{fig:cloud_diagram}, we show a simplified version of molecular clouds consisting of three layers: a dense core, a medium density layer and a lower density outer layer. In the lowest density layer, there are few collisions between H$_{2}$ and CO molecules. This leaves almost all CO molecules in the ground state. Therefore, only (1-0) absorption can be produced via the promotion of molecules to the first excited state. Few molecules will be in other states, so absorption from higher order lines is not produced. The medium density layer will have more collisional excitation occurring between H$_{2}$ and CO molecules, making CO(2-1) the dominant absorption line.

Additionally, due to the 115\,GHz interval in the CO(1-0) and CO(2-1) frequencies, different parts of the core may dominate at different frequencies due to spectral index variations or synchrotron opacity effects. This would produce a continuum which is smaller at higher frequencies.

On the left side of Figure \ref{fig:cloud_diagram}, we show a 230\,GHz continuum which passes through only the most diffuse parts of the obscuring cloud. Since all the molecules are in the ground state, no CO(2-1) absorption is produced. The slightly larger 115\,GHz continuum passes through a large volume of low density gas, leading to strong CO(1-0) absorption. The resulting CO(2-1)/CO(1-0) ratio is very small.

On the right side of Figure \ref{fig:cloud_diagram}, we show a 230\,GHz continuum which passes through a large volume of CO(2-1) thick gas. On the other hand, the volume of CO(1-0) thick gas passed through by the 115\,GHz continuum is significantly lower than in the first arrangement. Thus, there is a much larger velocity integrated optical depth ratio than the limit of 4 predicted by equation \ref{eq:opacityratio}. 

The combined effect of the molecular cloud structure and the varying size of the continuum source can therefore produce a wider range of CO(2-1)/CO(1-0) ratios than would be possible for homogeneous clouds which obscure a continuum whose size does not vary with frequency. It can also lead to apparently negative excitation temperatures. 



\section{Conclusions}

We present analysis of intrinsic molecular absorption lines -- those where the background radio continuum and the gas which absorbs it are within the same system. Our data includes many new CO, CN, HCN and HCO$^{+}$ absorption lines, as well as CO(2-1) emission. In these objects, the line's redshift unambiguously reveals the speed and direction of motion of the absorbing gas, allowing us to study its bulk motions. Our main conclusions are as follows:

\begin{itemize}
    \item In galaxies where our line of sight passes through an edge-on or close to edge-on disk of molecular gas, we find low velocity dispersion clouds (typically <9\,km/s) with no net inflow our outflow (Figures \ref{fig:Histogram} and \ref{fig:infall_vs_sigma}). In these respects, the clouds are similar those seen throughout of the Milky Way. This is to be expected, given that an edge-on alignment maximises our effective line of sight through the galaxy's disk. Therefore, the chance of the disk's molecular clouds aligning with the continuum is maximised. 
    \item When excluding disks, we find that absorption regions have significantly different properties. They have wide velocity dispersions (typically >9\,km/s), and in terms of their line of sight velocities, appear to split into two groups. One has red and blueshifted line of sight velocities between -100 and 100\,km/s. We argue that these clouds are likely within the circumnuclear disk surrounding the AGN and that their small line of sight velocities are a result of elliptical orbits about the galaxy centre (Figure \ref{fig:schematic_of_absorption_regions}). A second group has line of sight velocities of 150-550\,km/s, with no blueshifted counterpart. These exclusively positive velocities imply the bulk movement of molecular gas towards the galaxy centre. We interpret this as evidence of cold accretion onto the galactic centres, which is likely to contribute to AGN fuelling.
    \item We find a wide range in the CO(2-1)/CO(1-0) ratios of the absorption regions. In around half of cases, the ratio implies a non-physical negative excitation temperature. We show that the high CO(2-1)/CO(1-0) ratios can be explained by variation in the structure of the continuum source with frequency, combined with substructure in the molecular clouds.
\end{itemize}

\section*{Acknowledgements}

We thank Tom Oosterloo for providing the HI spectrum of NGC 6868 and Seiji Kameno for helpful discussions when analysing the ALMA data of NGC 1052. We also thank Jeff Mangum for helpful discussions on column densities. T.R. thanks the Waterloo Centre for Astrophysics and generous funding to B.R.M. from the Canadian Space Agency and the National Science and Engineering Research Council of Canada. A.C.E. acknowledges support from STFC grant ST/T000244/1. HRR acknowledges support from an STFC Ernest Rutherford Fellowship and an Anne McLaren Fellowship. P.S. acknowledges support by the ANR grant LYRICS (ANR-16-CE31-0011).

This paper makes use of the following ALMA data: 2017.1.00629.S, 2021.1.00766.S, 2018.1.01471, 2018.1.00581, 2013.1.00229, 2012.1.00988, 2011.0.00735, 2015.1.00591, 2015.1.01290, 2013.1.01225, 2016.1.00375, 2017.1.00301 and 2017.1.01638. ALMA is a partnership of ESO (representing its member states), NSF (USA) and NINS (Japan), together with NRC (Canada), MOST and ASIAA (Taiwan), and KASI (Republic of Korea), in cooperation with the Republic of Chile. The Joint ALMA Observatory is operated by ESO, AUI/NRAO and NAOJ. The National Radio Astronomy Observatory is a facility of the National Science Foundation operated under cooperative agreement by Associated Universities, Inc.

This research made use of \texttt{Astropy} \citep{the_astropy_collaboration_astropy_2013,the_astropy_collaboration_astropy_2018}, \texttt{Matplotlib} \citep{hunter_matplotlib_2007}, \texttt{numpy} \citep{walt_numpy_2011,harris_array_2020}, \texttt{Python} \citep{van_rossum_python_2009}, \texttt{Scipy} \citep{jones_scipy_2011,virtanen_scipy_2020} and \texttt{Aplpy} \citep[][]{aplpy}. We thank their developers for maintaining them and making them freely available.

\section*{Data Availability}
 
All ALMA data presented in this paper is publicly available from the NRAO's archive.


\bibliographystyle{mnras}

\begin{thebibliography}{}
\makeatletter
\relax
\def\mn@urlcharsother{\let\do\@makeother \do\$\do\&\do\#\do\^\do\_\do\%\do\~}
\def\mn@doi{\begingroup\mn@urlcharsother \@ifnextchar [ {\mn@doi@}
  {\mn@doi@[]}}
\def\mn@doi@[#1]#2{\def\@tempa{#1}\ifx\@tempa\@empty \href
  {http://dx.doi.org/#2} {doi:#2}\else \href {http://dx.doi.org/#2} {#1}\fi
  \endgroup}
\def\mn@eprint#1#2{\mn@eprint@#1:#2::\@nil}
\def\mn@eprint@arXiv#1{\href {http://arxiv.org/abs/#1} {{\tt arXiv:#1}}}
\def\mn@eprint@dblp#1{\href {http://dblp.uni-trier.de/rec/bibtex/#1.xml}
  {dblp:#1}}
\def\mn@eprint@#1:#2:#3:#4\@nil{\def\@tempa {#1}\def\@tempb {#2}\def\@tempc
  {#3}\ifx \@tempc \@empty \let \@tempc \@tempb \let \@tempb \@tempa \fi \ifx
  \@tempb \@empty \def\@tempb {arXiv}\fi \@ifundefined
  {mn@eprint@\@tempb}{\@tempb:\@tempc}{\expandafter \expandafter \csname
  mn@eprint@\@tempb\endcsname \expandafter{\@tempc}}}

\bibitem[\protect\citeauthoryear{{Alcorn} et~al.,}{{Alcorn}
  et~al.}{2023}]{Alcorn2023}
{Alcorn} L.~Y.,  et~al., 2023, \mn@doi [\mnras] {10.1093/mnras/stad948}, \href
  {https://ui.adsabs.harvard.edu/abs/2023MNRAS.tmp..939A} {}

\bibitem[\protect\citeauthoryear{Allen, Ettori  \& Fabian}{Allen
  et~al.}{2001}]{Allen2001}
Allen S.,  Ettori S.,   Fabian A.,  2001, \mn@doi [\mnras]
  {10.1046/j.1365-8711.2001.04318.x}, 324, 877

\bibitem[\protect\citeauthoryear{{Allison} et~al.,}{{Allison}
  et~al.}{2019}]{Allison2019}
{Allison} J.~R.,  et~al., 2019, \mn@doi [\mnras] {10.1093/mnras/sty2852}, \href
  {https://ui.adsabs.harvard.edu/abs/2019MNRAS.482.2934A} {482, 2934}

\bibitem[\protect\citeauthoryear{{Augusto}, {Edge}  \& {Chandler}}{{Augusto}
  et~al.}{2006}]{Augusto2006}
{Augusto} P.,  {Edge} A.~C.,   {Chandler} C.~J.,  2006, \mn@doi [\mnras]
  {10.1111/j.1365-2966.2005.09965.x}, \href
  {https://ui.adsabs.harvard.edu/abs/2006MNRAS.367..366A} {367, 366}

\bibitem[\protect\citeauthoryear{{Baek}, {Chung}, {Edge}, {Rose}, {Kim}  \&
  {Jung}}{{Baek} et~al.}{2022}]{Baek2022}
{Baek} J.,  {Chung} A.,  {Edge} A.,  {Rose} T.,  {Kim} J.-W.,   {Jung} T.,
  2022, \mn@doi [\apj] {10.3847/1538-4357/ac6de6}, \href
  {https://ui.adsabs.harvard.edu/abs/2022ApJ...932...64B} {932, 64}

\bibitem[\protect\citeauthoryear{{Baidoo}, {Perley}, {Eilek}, {Smirnov},
  {Vacca}  \& {En{\ss}lin}}{{Baidoo} et~al.}{2023}]{Baidoo2023}
{Baidoo} L.,  {Perley} R.~A.,  {Eilek} J.,  {Smirnov} O.,  {Vacca} V.,
  {En{\ss}lin} T.,  2023, \mn@doi [\apj] {10.3847/1538-4357/acebc5}, \href
  {https://ui.adsabs.harvard.edu/abs/2023ApJ...955...16B} {955, 16}

\bibitem[\protect\citeauthoryear{{Benedetti} et~al.,}{{Benedetti}
  et~al.}{2023}]{Benedetti2023}
{Benedetti} J. P.~V.,  et~al., 2023, \mn@doi [\mnras] {10.1093/mnras/stad1148},
  \href {https://ui.adsabs.harvard.edu/abs/2023MNRAS.522.2570B} {522, 2570}

\bibitem[\protect\citeauthoryear{{Bergin} \& {Tafalla}}{{Bergin} \&
  {Tafalla}}{2007}]{Bergin2007}
{Bergin} E.~A.,  {Tafalla} M.,  2007, \mn@doi [\araa]
  {10.1146/annurev.astro.45.071206.100404}, \href
  {https://ui.adsabs.harvard.edu/abs/2007ARA&A..45..339B} {45, 339}

\bibitem[\protect\citeauthoryear{{Bianchi}, {Matt}, {Fiore}, {Fabian},
  {Iwasawa}  \& {Nicastro}}{{Bianchi} et~al.}{2002}]{Bianchi2002}
{Bianchi} S.,  {Matt} G.,  {Fiore} F.,  {Fabian} A.~C.,  {Iwasawa} K.,
  {Nicastro} F.,  2002, \mn@doi [\aap] {10.1051/0004-6361:20021414}, \href
  {https://ui.adsabs.harvard.edu/abs/2002A&A...396..793B} {396, 793}

\bibitem[\protect\citeauthoryear{{B{\"o}hringer} et~al.,}{{B{\"o}hringer}
  et~al.}{2004}]{Bohringer2004}
{B{\"o}hringer} H.,  et~al., 2004, \mn@doi [\aap] {10.1051/0004-6361:20034484},
  \href {https://ui.adsabs.harvard.edu/abs/2004A%26A...425..367B} {425, 367}

\bibitem[\protect\citeauthoryear{{Boizelle} et~al.,}{{Boizelle}
  et~al.}{2021}]{Boizelle2021}
{Boizelle} B.~D.,  et~al., 2021, \mn@doi [\apj] {10.3847/1538-4357/abd24d},
  \href {https://ui.adsabs.harvard.edu/abs/2021ApJ...908...19B} {908, 19}

\bibitem[\protect\citeauthoryear{{Bolatto}, {Leroy}, {Israel}  \&
  {Jackson}}{{Bolatto} et~al.}{2003}]{Bolatto2003}
{Bolatto} A.~D.,  {Leroy} A.,  {Israel} F.~P.,   {Jackson} J.~M.,  2003,
  \mn@doi [\apj] {10.1086/377230}, \href
  {http://adsabs.harvard.edu/abs/2003ApJ...595..167B} {595, 167}

\bibitem[\protect\citeauthoryear{{Bolatto}, {Wolfire}  \& {Leroy}}{{Bolatto}
  et~al.}{2013}]{Bolatto2013}
{Bolatto} A.~D.,  {Wolfire} M.,   {Leroy} A.~K.,  2013, \mn@doi [\araa]
  {10.1146/annurev-astro-082812-140944}, \href
  {http://adsabs.harvard.edu/abs/2013ARA%26A..51..207B} {51, 207}

\bibitem[\protect\citeauthoryear{{Buote}, {Brighenti}  \& {Mathews}}{{Buote}
  et~al.}{2004}]{Buote2004}
{Buote} D.~A.,  {Brighenti} F.,   {Mathews} W.~G.,  2004, \mn@doi [\apjl]
  {10.1086/422097}, \href {http://adsabs.harvard.edu/abs/2004ApJ...607L..91B}
  {607, L91}

\bibitem[\protect\citeauthoryear{{Burke-Spolaor}, {Ekers}, {Massardi},
  {Murphy}, {Partridge}, {Ricci}  \& {Sadler}}{{Burke-Spolaor}
  et~al.}{2009}]{Burke2009}
{Burke-Spolaor} S.,  {Ekers} R.~D.,  {Massardi} M.,  {Murphy} T.,  {Partridge}
  B.,  {Ricci} R.,   {Sadler} E.~M.,  2009, \mn@doi [\mnras]
  {10.1111/j.1365-2966.2009.14556.x}, \href
  {https://ui.adsabs.harvard.edu/abs/2009MNRAS.395..504B} {395, 504}

\bibitem[\protect\citeauthoryear{{Choi}, {Leighly}, {Terndrup}, {Gallagher}  \&
  {Richards}}{{Choi} et~al.}{2020}]{Choi2020}
{Choi} H.,  {Leighly} K.~M.,  {Terndrup} D.~M.,  {Gallagher} S.~C.,
  {Richards} G.~T.,  2020, \mn@doi [\apj] {10.3847/1538-4357/ab6f72}, \href
  {https://ui.adsabs.harvard.edu/abs/2020ApJ...891...53C} {891, 53}

\bibitem[\protect\citeauthoryear{{Combes} \& {Gupta}}{{Combes} \&
  {Gupta}}{2024}]{Combes2024}
{Combes} F.,  {Gupta} N.,  2024, \mn@doi [\aap] {10.1051/0004-6361/202348386},
  \href {https://ui.adsabs.harvard.edu/abs/2024A&A...683A..20C} {683, A20}

\bibitem[\protect\citeauthoryear{{Crook}, {Huchra}, {Martimbeau}, {Masters},
  {Jarrett}  \& {Macri}}{{Crook} et~al.}{2007}]{Crook2007}
{Crook} A.~C.,  {Huchra} J.~P.,  {Martimbeau} N.,  {Masters} K.~L.,  {Jarrett}
  T.,   {Macri} L.~M.,  2007, \mn@doi [\apj] {10.1086/510201}, \href
  {https://ui.adsabs.harvard.edu/abs/2007ApJ...655..790C} {655, 790}

\bibitem[\protect\citeauthoryear{{David} et~al.,}{{David}
  et~al.}{2011}]{David2011}
{David} L.~P.,  et~al., 2011, \mn@doi [\apj] {10.1088/0004-637X/728/2/162},
  \href {http://adsabs.harvard.edu/abs/2011ApJ...728..162D} {728, 162}

\bibitem[\protect\citeauthoryear{David et~al.,}{David et~al.}{2014}]{David2014}
David L.~P.,  et~al., 2014, \apj, 792, 94

\bibitem[\protect\citeauthoryear{{Edge}, {Ivison}, {Smail}, {Blain}  \&
  {Kneib}}{{Edge} et~al.}{1999}]{Edge1999}
{Edge} A.~C.,  {Ivison} R.~J.,  {Smail} I.,  {Blain} A.~W.,   {Kneib} J.-P.,
  1999, \mn@doi [\mnras] {10.1046/j.1365-8711.1999.02539.x}, \href
  {http://adsabs.harvard.edu/abs/1999MNRAS.306..599E} {306, 599}

\bibitem[\protect\citeauthoryear{{Elmouttie}, {Krause}, {Haynes}  \&
  {Jones}}{{Elmouttie} et~al.}{1998}]{Elmouttie1998}
{Elmouttie} M.,  {Krause} M.,  {Haynes} R.~F.,   {Jones} K.~L.,  1998, \mn@doi
  [\mnras] {10.1046/j.1365-8711.1998.02002.x}, \href
  {https://ui.adsabs.harvard.edu/abs/1998MNRAS.300.1119E} {300, 1119}

\bibitem[\protect\citeauthoryear{{Emonts} et~al.,}{{Emonts}
  et~al.}{2024}]{Emonts2024}
{Emonts} B. H.~C.,  et~al., 2024, \mn@doi [\apj] {10.3847/1538-4357/ad198d},
  \href {https://ui.adsabs.harvard.edu/abs/2024ApJ...962..187E} {962, 187}

\bibitem[\protect\citeauthoryear{{Esquej} et~al.,}{{Esquej}
  et~al.}{2014}]{Esquej2014}
{Esquej} P.,  et~al., 2014, \mn@doi [\apj] {10.1088/0004-637X/780/1/86}, \href
  {https://ui.adsabs.harvard.edu/abs/2014ApJ...780...86E} {780, 86}

\bibitem[\protect\citeauthoryear{{Fabian}, {Ferland}, {Sanders}, {McNamara},
  {Pinto}  \& {Walker}}{{Fabian} et~al.}{2022}]{Fabian2022}
{Fabian} A.~C.,  {Ferland} G.~J.,  {Sanders} J.~S.,  {McNamara} B.~R.,  {Pinto}
  C.,   {Walker} S.~A.,  2022, \mn@doi [\mnras] {10.1093/mnras/stac2003}, \href
  {https://ui.adsabs.harvard.edu/abs/2022MNRAS.515.3336F} {515, 3336}

\bibitem[\protect\citeauthoryear{{Ferrarese}, {Ford}  \& {Jaffe}}{{Ferrarese}
  et~al.}{1996}]{Ferrarese1996}
{Ferrarese} L.,  {Ford} H.~C.,   {Jaffe} W.,  1996, \mn@doi [\apj]
  {10.1086/177876}, \href
  {https://ui.adsabs.harvard.edu/abs/1996ApJ...470..444F} {470, 444}

\bibitem[\protect\citeauthoryear{Gaspari, Melioli, Brighenti  \&
  D'Ercole}{Gaspari et~al.}{2011}]{Gaspari2010}
Gaspari M.,  Melioli C.,  Brighenti F.,   D'Ercole A.,  2011, \mn@doi [\mnras]
  {10.1111/j.1365-2966.2010.17688.x}, 411, 349

\bibitem[\protect\citeauthoryear{{Gaspari}, {Brighenti}  \& {Temi}}{{Gaspari}
  et~al.}{2015}]{Gaspari2015}
{Gaspari} M.,  {Brighenti} F.,   {Temi} P.,  2015, \mn@doi [\aap]
  {10.1051/0004-6361/201526151}, \href
  {http://adsabs.harvard.edu/abs/2015A%26A...579A..62G} {579, A62}

\bibitem[\protect\citeauthoryear{Gaspari et~al.}{Gaspari
  et~al.}{2018}]{Gaspari2018}
Gaspari M.,  et~al., 2018, \mn@doi [Astrophys. J.] {10.3847/1538-4357/aaaa1b},
  854, 167

\bibitem[\protect\citeauthoryear{Gastaldello et~al.}{Gastaldello
  et~al.}{2013}]{Gastaldello2013}
Gastaldello F.,  et~al., 2013, \mn@doi [Astrophys. J.]
  {10.1088/0004-637X/770/1/56}, 770, 56

\bibitem[\protect\citeauthoryear{{Gavazzi}, {Zaccardo}, {Sanvito}, {Boselli}
  \& {Bonfanti}}{{Gavazzi} et~al.}{2004}]{Gavazzi2004}
{Gavazzi} G.,  {Zaccardo} A.,  {Sanvito} G.,  {Boselli} A.,   {Bonfanti} C.,
  2004, \mn@doi [\aap] {10.1051/0004-6361:20034105}, \href
  {https://ui.adsabs.harvard.edu/abs/2004A&A...417..499G} {417, 499}

\bibitem[\protect\citeauthoryear{{Godard}, {Falgarone}, {Gerin}, {Hily-Blant}
  \& {de Luca}}{{Godard} et~al.}{2010}]{Godard2010}
{Godard} B.,  {Falgarone} E.,  {Gerin} M.,  {Hily-Blant} P.,   {de Luca} M.,
  2010, \mn@doi [\aap] {10.1051/0004-6361/201014283}, \href
  {http://adsabs.harvard.edu/abs/2010A%26A...520A..20G} {520, A20}

\bibitem[\protect\citeauthoryear{{Gong} et~al.,}{{Gong}
  et~al.}{2016}]{Gong2016}
{Gong} Y.,  et~al., 2016, \mn@doi [\aap] {10.1051/0004-6361/201527334}, \href
  {https://ui.adsabs.harvard.edu/abs/2016A&A...588A.104G} {588, A104}

\bibitem[\protect\citeauthoryear{{Hamer} et~al.,}{{Hamer}
  et~al.}{2014}]{Hamer2014}
{Hamer} S.~L.,  et~al., 2014, \mn@doi [\mnras] {10.1093/mnras/stt1949}, \href
  {https://ui.adsabs.harvard.edu/abs/2014MNRAS.437..862H} {437, 862}

\bibitem[\protect\citeauthoryear{Hamer et~al.}{Hamer et~al.}{2016}]{Hamer2016}
Hamer S.~L.,  et~al., 2016, \mn@doi [\mnras] {10.1093/mnras/stw1054}, 460, 1758

\bibitem[\protect\citeauthoryear{Harris et~al.,}{Harris
  et~al.}{2020}]{harris_array_2020}
Harris C.~R.,  et~al., 2020, \mn@doi [Nature] {10.1038/s41586-020-2649-2}, 585,
  357

\bibitem[\protect\citeauthoryear{{Harrison}, {Costa}, {Tadhunter},
  {Fl{\"u}tsch}, {Kakkad}, {Perna}  \& {Vietri}}{{Harrison}
  et~al.}{2018}]{Harrison2018}
{Harrison} C.~M.,  {Costa} T.,  {Tadhunter} C.~N.,  {Fl{\"u}tsch} A.,  {Kakkad}
  D.,  {Perna} M.,   {Vietri} G.,  2018, \mn@doi [Nature Astronomy]
  {10.1038/s41550-018-0403-6}, \href
  {https://ui.adsabs.harvard.edu/abs/2018NatAs...2..198H} {2, 198}

\bibitem[\protect\citeauthoryear{{Hern{\'a}ndez}, {Ghosh}, {Salter}  \&
  {Momjian}}{{Hern{\'a}ndez} et~al.}{2008}]{Hernandez2008}
{Hern{\'a}ndez} H.,  {Ghosh} T.,  {Salter} C.~J.,   {Momjian} E.,  2008, in
  {Minchin} R.,  {Momjian} E.,  eds,  American Institute of Physics Conference
  Series Vol. 1035, The Evolution of Galaxies Through the Neutral Hydrogen
  Window. pp 214--217, \mn@doi{10.1063/1.2973584}

\bibitem[\protect\citeauthoryear{Hogan}{Hogan}{2014}]{Hogan_thesis}
Hogan M.,  2014, \url
  {http://etheses.dur.ac.uk/11008/1/hogan_thesis.pdf?DDD25+}

\bibitem[\protect\citeauthoryear{{Hogan} et~al.,}{{Hogan}
  et~al.}{2015a}]{Hogan2015b}
{Hogan} M.~T.,  et~al., 2015a, \mn@doi [\mnras] {10.1093/mnras/stv1517}, \href
  {http://adsabs.harvard.edu/abs/2015MNRAS.453.1201H} {453, 1201}

\bibitem[\protect\citeauthoryear{Hogan et~al.,}{Hogan
  et~al.}{2015b}]{Hogan2015}
Hogan M.~T.,  et~al., 2015b, \mn@doi [\mnras] {10.1093/mnras/stv1518}, 453,
  1223

\bibitem[\protect\citeauthoryear{Hunter}{Hunter}{2007}]{hunter_matplotlib_2007}
Hunter J.~D.,  2007, \mn@doi [Comput. Sci. Eng.] {10.1109/MCSE.2007.55}, 9, 90

\bibitem[\protect\citeauthoryear{{Israel}, {van Dishoeck}, {Baas}, {Koornneef},
  {Black}  \& {de Graauw}}{{Israel} et~al.}{1990}]{Israel1990}
{Israel} F.~P.,  {van Dishoeck} E.~F.,  {Baas} F.,  {Koornneef} J.,  {Black}
  J.~H.,   {de Graauw} T.,  1990, \aap, \href
  {https://ui.adsabs.harvard.edu/abs/1990A&A...227..342I} {227, 342}

\bibitem[\protect\citeauthoryear{{Izumi}, {Wada}, {Fukushige}, {Hamamura}  \&
  {Kohno}}{{Izumi} et~al.}{2018}]{Izumi2018}
{Izumi} T.,  {Wada} K.,  {Fukushige} R.,  {Hamamura} S.,   {Kohno} K.,  2018,
  \mn@doi [\apj] {10.3847/1538-4357/aae20b}, \href
  {https://ui.adsabs.harvard.edu/abs/2018ApJ...867...48I} {867, 48}

\bibitem[\protect\citeauthoryear{{Izumi} et~al.,}{{Izumi}
  et~al.}{2023}]{Izumi2023}
{Izumi} T.,  et~al., 2023, \mn@doi [Science] {10.1126/science.adf0569}, \href
  {https://ui.adsabs.harvard.edu/abs/2023Sci...382..554I} {382, 554}

\bibitem[\protect\citeauthoryear{{Jaffe} \& {McNamara}}{{Jaffe} \&
  {McNamara}}{1994}]{Jaffe1994}
{Jaffe} W.,  {McNamara} B.~R.,  1994, \mn@doi [\apj] {10.1086/174708}, \href
  {https://ui.adsabs.harvard.edu/abs/1994ApJ...434..110J} {434, 110}

\bibitem[\protect\citeauthoryear{Johnson, Markevitch, Wegner, Jones  \&
  Forman}{Johnson et~al.}{2010}]{Johnson2010}
Johnson R.~E.,  Markevitch M.,  Wegner G.~A.,  Jones C.,   Forman W.~R.,  2010,
  \mn@doi [\apj] {10.1088/0004-637x/710/2/1776}, 710, 1776

\bibitem[\protect\citeauthoryear{Jones, Oliphant  \& Peterson}{Jones
  et~al.}{2011}]{jones_scipy_2011}
Jones E.,  Oliphant T.,   Peterson P.,  2011, {{SciPy Open}} Source Scientific
  Tools for {{Python}}, \url {www.scipy.org}

\bibitem[\protect\citeauthoryear{{Kameno} et~al.,}{{Kameno}
  et~al.}{2020}]{Kameno2020}
{Kameno} S.,  et~al., 2020, \mn@doi [\apj] {10.3847/1538-4357/ab8bd6}, \href
  {https://ui.adsabs.harvard.edu/abs/2020ApJ...895...73K} {895, 73}

\bibitem[\protect\citeauthoryear{{Kameno}, {Harikane}, {Sawada-Satoh},
  {Sawada}, {Saito}, {Nakanishi}, {Humphreys}  \& {Impellizzeri}}{{Kameno}
  et~al.}{2024}]{Kameno2024}
{Kameno} S.,  {Harikane} Y.,  {Sawada-Satoh} S.,  {Sawada} T.,  {Saito} T.,
  {Nakanishi} K.,  {Humphreys} E.,   {Impellizzeri} C.~M.~V.,  2024, \mn@doi
  [arXiv e-prints] {10.48550/arXiv.2402.06166}, \href
  {https://ui.adsabs.harvard.edu/abs/2024arXiv240206166K} {p. arXiv:2402.06166}

\bibitem[\protect\citeauthoryear{{Laha}, {Reynolds}, {Reeves}, {Kriss},
  {Guainazzi}, {Smith}, {Veilleux}  \& {Proga}}{{Laha} et~al.}{2021}]{Laha2021}
{Laha} S.,  {Reynolds} C.~S.,  {Reeves} J.,  {Kriss} G.,  {Guainazzi} M.,
  {Smith} R.,  {Veilleux} S.,   {Proga} D.,  2021, \mn@doi [Nature Astronomy]
  {10.1038/s41550-020-01255-2}, \href
  {https://ui.adsabs.harvard.edu/abs/2021NatAs...5...13L} {5, 13}

\bibitem[\protect\citeauthoryear{{Lane}, {Clarke}, {Taylor}, {Perley}  \&
  {Kassim}}{{Lane} et~al.}{2004}]{Lane2004}
{Lane} W.~M.,  {Clarke} T.~E.,  {Taylor} G.~B.,  {Perley} R.~A.,   {Kassim}
  N.~E.,  2004, \mn@doi [\aj] {10.1086/379858}, \href
  {https://ui.adsabs.harvard.edu/abs/2004AJ....127...48L} {127, 48}

\bibitem[\protect\citeauthoryear{{Lauer} et~al.,}{{Lauer}
  et~al.}{2005}]{Lauer2005}
{Lauer} T.~R.,  et~al., 2005, \mn@doi [\aj] {10.1086/429565}, \href
  {https://ui.adsabs.harvard.edu/abs/2005AJ....129.2138L} {129, 2138}

\bibitem[\protect\citeauthoryear{{Mangum} \& {Shirley}}{{Mangum} \&
  {Shirley}}{2015}]{Magnum2015}
{Mangum} J.~G.,  {Shirley} Y.~L.,  2015, \mn@doi [\pasp] {10.1086/680323},
  \href {http://adsabs.harvard.edu/abs/2015PASP..127..266M} {127, 266}

\bibitem[\protect\citeauthoryear{{Matsumoto}, {Fukazawa}, {Nakazawa}, {Iyomoto}
   \& {Makishima}}{{Matsumoto} et~al.}{2001}]{Matsumoto2001}
{Matsumoto} Y.,  {Fukazawa} Y.,  {Nakazawa} K.,  {Iyomoto} N.,   {Makishima}
  K.,  2001, \mn@doi [\pasj] {10.1093/pasj/53.3.475}, \href
  {https://ui.adsabs.harvard.edu/abs/2001PASJ...53..475M} {53, 475}

\bibitem[\protect\citeauthoryear{{McMullin}, {Waters}, {Schiebel}, {Young}  \&
  {Golap}}{{McMullin} et~al.}{2007}]{CASA}
{McMullin} J.~P.,  {Waters} B.,  {Schiebel} D.,  {Young} W.,   {Golap} K.,
  2007, in {Shaw} R.~A.,  {Hill} F.,   {Bell} D.~J.,  eds,  Astronomical
  Society of the Pacific Conference Series Vol. 376, Astronomical Data Analysis
  Software and Systems XVI. p.~127

\bibitem[\protect\citeauthoryear{{McNamara} et~al.,}{{McNamara}
  et~al.}{2000}]{McNamara2000}
{McNamara} B.~R.,  et~al., 2000, \mn@doi [\apjl] {10.1086/312662}, \href
  {http://adsabs.harvard.edu/abs/2000ApJ...534L.135M} {534, L135}

\bibitem[\protect\citeauthoryear{McNamara et~al.,}{McNamara
  et~al.}{2001}]{McNamara2001}
McNamara B.~R.,  et~al., 2001, \mn@doi [The Astrophysical Journal]
  {10.1086/338326}, 562, L149

\bibitem[\protect\citeauthoryear{McNamara, Russell, Nulsen, Hogan, Fabian,
  Pulido  \& Edge}{McNamara et~al.}{2016}]{McNamara2016}
McNamara B.~R.,  Russell H.~R.,  Nulsen P. E.~J.,  Hogan M.~T.,  Fabian A.~C.,
  Pulido F.,   Edge A.~C.,  2016, \apj, 830, 79

\bibitem[\protect\citeauthoryear{{Morganti} et~al.,}{{Morganti}
  et~al.}{2023}]{Morganti2023}
{Morganti} R.,  et~al., 2023, \mn@doi [\aap] {10.1051/0004-6361/202347117},
  \href {https://ui.adsabs.harvard.edu/abs/2023A&A...678A..42M} {678, A42}

\bibitem[\protect\citeauthoryear{{M{\"u}ller S{\'a}nchez}, {Davies},
  {Eisenhauer}, {Tacconi}, {Genzel}  \& {Sternberg}}{{M{\"u}ller S{\'a}nchez}
  et~al.}{2006}]{MullerSanchez2006}
{M{\"u}ller S{\'a}nchez} F.,  {Davies} R.~I.,  {Eisenhauer} F.,  {Tacconi}
  L.~J.,  {Genzel} R.,   {Sternberg} A.,  2006, \mn@doi [\aap]
  {10.1051/0004-6361:20054387}, \href
  {https://ui.adsabs.harvard.edu/abs/2006A&A...454..481M} {454, 481}

\bibitem[\protect\citeauthoryear{{Nagai} et~al.,}{{Nagai}
  et~al.}{2019}]{Nagai19}
{Nagai} H.,  et~al., 2019, \mn@doi [\apj] {10.3847/1538-4357/ab3e6e}, \href
  {https://ui.adsabs.harvard.edu/abs/2019ApJ...883..193N} {883, 193}

\bibitem[\protect\citeauthoryear{{Nakahara}, {Doi}, {Murata}, {Nakamura},
  {Hada}, {Asada}, {Sawada-Satoh}  \& {Kameno}}{{Nakahara}
  et~al.}{2020}]{Nakahara2020}
{Nakahara} S.,  {Doi} A.,  {Murata} Y.,  {Nakamura} M.,  {Hada} K.,  {Asada}
  K.,  {Sawada-Satoh} S.,   {Kameno} S.,  2020, \mn@doi [\aj]
  {10.3847/1538-3881/ab465b}, \href
  {https://ui.adsabs.harvard.edu/abs/2020AJ....159...14N} {159, 14}

\bibitem[\protect\citeauthoryear{{O'Dea}, {Baum}, {Maloney}, {Tacconi}  \&
  {Sparks}}{{O'Dea} et~al.}{1994}]{ODea1994}
{O'Dea} C.~P.,  {Baum} S.~A.,  {Maloney} P.~R.,  {Tacconi} L.~J.,   {Sparks}
  W.~B.,  1994, \mn@doi [\apj] {10.1086/173742}, \href
  {http://adsabs.harvard.edu/abs/1994ApJ...422..467O} {422, 467}

\bibitem[\protect\citeauthoryear{{Oka}, {Nagai}, {Kamegai}  \& {Tanaka}}{{Oka}
  et~al.}{2011}]{Oka2011}
{Oka} T.,  {Nagai} M.,  {Kamegai} K.,   {Tanaka} K.,  2011, \mn@doi [\apj]
  {10.1088/0004-637X/732/2/120}, \href
  {https://ui.adsabs.harvard.edu/abs/2011ApJ...732..120O} {732, 120}

\bibitem[\protect\citeauthoryear{{Olivares} et~al.,}{{Olivares}
  et~al.}{2019}]{Olivares2019}
{Olivares} V.,  et~al., 2019, arXiv e-prints, \href
  {https://ui.adsabs.harvard.edu/abs/2019arXiv190209164O} {}

\bibitem[\protect\citeauthoryear{{Oosterloo}, {Morganti}  \&
  {Murthy}}{{Oosterloo} et~al.}{2023}]{Oosterloo2023}
{Oosterloo} T.,  {Morganti} R.,   {Murthy} S.,  2023, \mn@doi [Nature
  Astronomy] {10.1038/s41550-023-02138-y}, \href
  {https://ui.adsabs.harvard.edu/abs/2023NatAs.tmp..254O} {}

\bibitem[\protect\citeauthoryear{{Parikh}, {Saglia}, {Thomas}, {Mehrgan},
  {Bender}  \& {Maraston}}{{Parikh} et~al.}{2024}]{Parikh2024}
{Parikh} T.,  {Saglia} R.,  {Thomas} J.,  {Mehrgan} K.,  {Bender} R.,
  {Maraston} C.,  2024, \mn@doi [arXiv e-prints] {10.48550/arXiv.2402.06628},
  \href {https://ui.adsabs.harvard.edu/abs/2024arXiv240206628P} {p.
  arXiv:2402.06628}

\bibitem[\protect\citeauthoryear{{Parkin} et~al.,}{{Parkin}
  et~al.}{2012}]{Parkin2012}
{Parkin} T.~J.,  et~al., 2012, \mn@doi [\mnras]
  {10.1111/j.1365-2966.2012.20778.x}, \href
  {https://ui.adsabs.harvard.edu/abs/2012MNRAS.422.2291P} {422, 2291}

\bibitem[\protect\citeauthoryear{{Pizzolato} \& {Soker}}{{Pizzolato} \&
  {Soker}}{2005}]{Pizzolato2005}
{Pizzolato} F.,  {Soker} N.,  2005, \mn@doi [\apj] {10.1086/444344}, \href
  {http://adsabs.harvard.edu/abs/2005ApJ...632..821P} {632, 821}

\bibitem[\protect\citeauthoryear{{Rickes}, {Pastoriza}  \& {Bonatto}}{{Rickes}
  et~al.}{2008}]{Rickes2008}
{Rickes} M.~G.,  {Pastoriza} M.~G.,   {Bonatto} C.,  2008, \mn@doi [\mnras]
  {10.1111/j.1365-2966.2007.12724.x}, \href
  {https://ui.adsabs.harvard.edu/abs/2008MNRAS.384.1427R} {384, 1427}

\bibitem[\protect\citeauthoryear{{Robitaille} \& {Bressert}}{{Robitaille} \&
  {Bressert}}{2012}]{aplpy}
{Robitaille} T.,  {Bressert} E.,  2012, {APLpy: Astronomical Plotting Library
  in Python}, Astrophysics Source Code Library (\mn@eprint {ascl} {1208.017})

\bibitem[\protect\citeauthoryear{{Roman-Duval}, {Jackson}, {Heyer}, {Rathborne}
   \& {Simon}}{{Roman-Duval} et~al.}{2010}]{RomanDuval2010}
{Roman-Duval} J.,  {Jackson} J.~M.,  {Heyer} M.,  {Rathborne} J.,   {Simon} R.,
   2010, \mn@doi [\apj] {10.1088/0004-637X/723/1/492}, \href
  {https://ui.adsabs.harvard.edu/abs/2010ApJ...723..492R} {723, 492}

\bibitem[\protect\citeauthoryear{{Rose} et~al.,}{{Rose}
  et~al.}{2019a}]{Rose2019a}
{Rose} T.,  et~al., 2019a, \mn@doi [\mnras] {10.1093/mnras/stz406}, \href
  {https://ui.adsabs.harvard.edu/abs/2019MNRAS.485..229R} {485, 229}

\bibitem[\protect\citeauthoryear{Rose et~al.,}{Rose et~al.}{2019b}]{Rose2019b}
Rose T.,  et~al., 2019b, \mn@doi [\mnras] {10.1093/mnras/stz2138}, 489, 349

\bibitem[\protect\citeauthoryear{{Rose} et~al.,}{{Rose}
  et~al.}{2020}]{Rose2020}
{Rose} T.,  et~al., 2020, \mn@doi [\mnras] {10.1093/mnras/staa1474}, \href
  {https://ui.adsabs.harvard.edu/abs/2020MNRAS.496..364R} {496, 364}

\bibitem[\protect\citeauthoryear{{Rose} et~al.,}{{Rose}
  et~al.}{2022}]{Rose2022}
{Rose} T.,  et~al., 2022, \mn@doi [\mnras] {10.1093/mnras/stab3217}, \href
  {https://ui.adsabs.harvard.edu/abs/2022MNRAS.509.2869R} {509, 2869}

\bibitem[\protect\citeauthoryear{{Rose} et~al.,}{{Rose}
  et~al.}{2023}]{Rose2023}
{Rose} T.,  et~al., 2023, \mn@doi [MNRAS] {10.1093/mnras/stac3194}, \href
  {https://ui.adsabs.harvard.edu/abs/2023MNRAS.518..878R} {518, 878}

\bibitem[\protect\citeauthoryear{{Rose} et~al.,}{{Rose}
  et~al.}{2024}]{Rose2024}
{Rose} T.,  et~al., 2024, \mn@doi [\mnras] {10.1093/mnras/stae213}, \href
  {https://ui.adsabs.harvard.edu/abs/2024MNRAS.528.3441R} {528, 3441}

\bibitem[\protect\citeauthoryear{Ruffa et~al.,}{Ruffa et~al.}{2019}]{Ruffa2019}
Ruffa I.,  et~al., 2019, \mn@doi [\mnras] {10.1093/mnras/stz255}, 484, 4239

\bibitem[\protect\citeauthoryear{{Ruffa} et~al.,}{{Ruffa}
  et~al.}{2023}]{Ruffa2023}
{Ruffa} I.,  et~al., 2023, \mn@doi [\mnras] {10.1093/mnras/stad1119}, \href
  {https://ui.adsabs.harvard.edu/abs/2023MNRAS.522.6170R} {522, 6170}

\bibitem[\protect\citeauthoryear{{Savini} et~al.,}{{Savini}
  et~al.}{2019}]{Savini2019}
{Savini} F.,  et~al., 2019, \mn@doi [\aap] {10.1051/0004-6361/201833882}, \href
  {https://ui.adsabs.harvard.edu/abs/2019A&A...622A..24S} {622, A24}

\bibitem[\protect\citeauthoryear{{Sawada-Satoh}, {Kameno}  \&
  {Trippe}}{{Sawada-Satoh} et~al.}{2022}]{Swada-Satoh2022}
{Sawada-Satoh} S.,  {Kameno} S.,   {Trippe} S.,  2022, \mn@doi [\aap]
  {10.1051/0004-6361/202244047}, \href
  {https://ui.adsabs.harvard.edu/abs/2022A&A...664L..11S} {664, L11}

\bibitem[\protect\citeauthoryear{{Schellenberger} et~al.,}{{Schellenberger}
  et~al.}{2020}]{Schellenberger2020}
{Schellenberger} G.,  et~al., 2020, \mn@doi [\apj] {10.3847/1538-4357/ab879c},
  \href {https://ui.adsabs.harvard.edu/abs/2020ApJ...894...72S} {894, 72}

\bibitem[\protect\citeauthoryear{{Singha}, {O'Dea}, {Gordon}, {Lawlor-Forsyth}
  \& {Baum}}{{Singha} et~al.}{2021}]{Singha2021}
{Singha} M.,  {O'Dea} C.~P.,  {Gordon} Y.~A.,  {Lawlor-Forsyth} C.,   {Baum}
  S.~A.,  2021, \mn@doi [\apj] {10.3847/1538-4357/ac06c7}, \href
  {https://ui.adsabs.harvard.edu/abs/2021ApJ...918...65S} {918, 65}

\bibitem[\protect\citeauthoryear{{Solanki}, {Ressler}, {Murchikova}, {Stone}
  \& {Morris}}{{Solanki} et~al.}{2023}]{Solanki2023}
{Solanki} S.,  {Ressler} S.~M.,  {Murchikova} L.,  {Stone} J.~M.,   {Morris}
  M.~R.,  2023, \mn@doi [\apj] {10.3847/1538-4357/acdb6f}, \href
  {https://ui.adsabs.harvard.edu/abs/2023ApJ...953...22S} {953, 22}

\bibitem[\protect\citeauthoryear{{Taylor}}{{Taylor}}{1996}]{Taylor1996}
{Taylor} G.~B.,  1996, \mn@doi [\apj] {10.1086/177874}, \href
  {http://adsabs.harvard.edu/abs/1996ApJ...470..394T} {470, 394}

\bibitem[\protect\citeauthoryear{{Taylor}, {Perley}, {Inoue}, {Kato}, {Tabara}
  \& {Aizu}}{{Taylor} et~al.}{1990}]{Taylor1990}
{Taylor} G.~B.,  {Perley} R.~A.,  {Inoue} M.,  {Kato} T.,  {Tabara} H.,
  {Aizu} K.,  1990, \mn@doi [\apj] {10.1086/169094}, \href
  {http://adsabs.harvard.edu/abs/1990ApJ...360...41T} {360, 41}

\bibitem[\protect\citeauthoryear{Temi, Amblard, Gitti, Brighenti, Gaspari,
  Mathews  \& David}{Temi et~al.}{2018}]{Temi2018}
Temi P.,  Amblard A.,  Gitti M.,  Brighenti F.,  Gaspari M.,  Mathews W.~G.,
  David L.,  2018, \apj, 858, 17

\bibitem[\protect\citeauthoryear{{The Astropy Collaboration} et~al.,}{{The
  Astropy Collaboration} et~al.}{2013}]{the_astropy_collaboration_astropy_2013}
{The Astropy Collaboration} et~al., 2013, \mn@doi [A\&A]
  {10.1051/0004-6361/201322068}, 558, A33

\bibitem[\protect\citeauthoryear{{The Astropy Collaboration} et~al.,}{{The
  Astropy Collaboration} et~al.}{2018}]{the_astropy_collaboration_astropy_2018}
{The Astropy Collaboration} et~al., 2018, \mn@doi [AJ]
  {10.3847/1538-3881/aabc4f}, 156, 123

\bibitem[\protect\citeauthoryear{{Tremblay} et~al.,}{{Tremblay}
  et~al.}{2012}]{Tremblay2012}
{Tremblay} G.~R.,  et~al., 2012, \mn@doi [\mnras]
  {10.1111/j.1365-2966.2012.21278.x}, \href
  {https://ui.adsabs.harvard.edu/abs/2012MNRAS.424.1042T} {424, 1042}

\bibitem[\protect\citeauthoryear{Tremblay et~al.}{Tremblay
  et~al.}{2016}]{Tremblay2016}
Tremblay G.~R.,  et~al., 2016, \mn@doi [Nature] {10.1038/nature17969}, 534, 218

\bibitem[\protect\citeauthoryear{{Tremblay} et~al.,}{{Tremblay}
  et~al.}{2018}]{Tremblay2018}
{Tremblay} G.~R.,  et~al., 2018, \mn@doi [\apj] {10.3847/1538-4357/aad6dd},
  \href {http://adsabs.harvard.edu/abs/2018ApJ...865...13T} {865, 13}

\bibitem[\protect\citeauthoryear{{Tully}, {Pierce}, {Huang}, {Saunders},
  {Verheijen}  \& {Witchalls}}{{Tully} et~al.}{1998}]{Tully1998}
{Tully} R.~B.,  {Pierce} M.~J.,  {Huang} J.-S.,  {Saunders} W.,  {Verheijen} M.
  A.~W.,   {Witchalls} P.~L.,  1998, \mn@doi [\aj] {10.1086/300379}, \href
  {https://ui.adsabs.harvard.edu/abs/1998AJ....115.2264T} {115, 2264}

\bibitem[\protect\citeauthoryear{{Tully}, {Rizzi}, {Shaya}, {Courtois},
  {Makarov}  \& {Jacobs}}{{Tully} et~al.}{2009}]{Tully2009}
{Tully} R.~B.,  {Rizzi} L.,  {Shaya} E.~J.,  {Courtois} H.~M.,  {Makarov}
  D.~I.,   {Jacobs} B.~A.,  2009, \mn@doi [\aj] {10.1088/0004-6256/138/2/323},
  \href {https://ui.adsabs.harvard.edu/abs/2009AJ....138..323T} {138, 323}

\bibitem[\protect\citeauthoryear{{Ursini} et~al.,}{{Ursini}
  et~al.}{2023}]{Ursini2023}
{Ursini} F.,  et~al., 2023, \mn@doi [\mnras] {10.1093/mnras/stac3189}, \href
  {https://ui.adsabs.harvard.edu/abs/2023MNRAS.519...50U} {519, 50}

\bibitem[\protect\citeauthoryear{Van~Rossum \& Drake}{Van~Rossum \&
  Drake}{2009}]{van_rossum_python_2009}
Van~Rossum G.,  Drake F.~L.,  2009, Python 3 {{Reference Manual}}.
{CreateSpace}, {Scotts Valley, CA}

\bibitem[\protect\citeauthoryear{{Vantyghem} et~al.,}{{Vantyghem}
  et~al.}{2017}]{Vantyghem2017}
{Vantyghem} A.~N.,  et~al., 2017, \mn@doi [\apj] {10.3847/1538-4357/aa8fd0},
  \href {https://ui.adsabs.harvard.edu/abs/2017ApJ...848..101V} {848, 101}

\bibitem[\protect\citeauthoryear{{Veilleux} \& {Bland-Hawthorn}}{{Veilleux} \&
  {Bland-Hawthorn}}{1997}]{Veilleux1997}
{Veilleux} S.,  {Bland-Hawthorn} J.,  1997, \mn@doi [\apjl] {10.1086/310588},
  \href {https://ui.adsabs.harvard.edu/abs/1997ApJ...479L.105V} {479, L105}

\bibitem[\protect\citeauthoryear{Virtanen et~al.,}{Virtanen
  et~al.}{2020}]{virtanen_scipy_2020}
Virtanen P.,  et~al., 2020, \mn@doi [Nature Methods]
  {10.1038/s41592-019-0686-2}, 17, 261

\bibitem[\protect\citeauthoryear{{Vollmer}, {Beckert}  \& {Duschl}}{{Vollmer}
  et~al.}{2004}]{Vollmer2004}
{Vollmer} B.,  {Beckert} T.,   {Duschl} W.~J.,  2004, \mn@doi [\aap]
  {10.1051/0004-6361:20034201}, \href
  {https://ui.adsabs.harvard.edu/abs/2004A&A...413..949V} {413, 949}

\bibitem[\protect\citeauthoryear{{Wiklind} \& {Combes}}{{Wiklind} \&
  {Combes}}{1996}]{Wiklind1996a}
{Wiklind} T.,  {Combes} F.,  1996, \mn@doi [\nat] {10.1038/379139a0}, \href
  {https://ui.adsabs.harvard.edu/abs/1996Natur.379..139W} {379, 139}

\bibitem[\protect\citeauthoryear{{Wiklind} \& {Combes}}{{Wiklind} \&
  {Combes}}{1997a}]{Wiklind1997a}
{Wiklind} T.,  {Combes} F.,  1997a, \aap, \href
  {https://ui.adsabs.harvard.edu/abs/1997A&A...324...51W} {324, 51}

\bibitem[\protect\citeauthoryear{{Wiklind} \& {Combes}}{{Wiklind} \&
  {Combes}}{1997b}]{Wiklind1997b}
{Wiklind} T.,  {Combes} F.,  1997b, \aap, \href
  {https://ui.adsabs.harvard.edu/abs/1997A&A...328...48W} {328, 48}

\bibitem[\protect\citeauthoryear{{Wild}, {Eckart}  \& {Wiklind}}{{Wild}
  et~al.}{1997}]{Wild1997}
{Wild} W.,  {Eckart} A.,   {Wiklind} T.,  1997, \aap, \href
  {https://ui.adsabs.harvard.edu/abs/1997A&A...322..419W} {322, 419}

\bibitem[\protect\citeauthoryear{Wise, McNamara, Nulsen, Houck  \& David}{Wise
  et~al.}{2007}]{Wise2006}
Wise M.~W.,  McNamara B.~R.,  Nulsen P. E.~J.,  Houck J.~C.,   David L.~P.,
  2007, \mn@doi [Astrophys. J.] {10.1086/512767}, 659, 1153

\bibitem[\protect\citeauthoryear{{Woo} \& {Urry}}{{Woo} \&
  {Urry}}{2002}]{Woo2002}
{Woo} J.-H.,  {Urry} C.~M.,  2002, \mn@doi [\apj] {10.1086/342878}, \href
  {https://ui.adsabs.harvard.edu/abs/2002ApJ...579..530W} {579, 530}

\bibitem[\protect\citeauthoryear{{de Vaucouleurs}, {de Vaucouleurs}, {Corwin},
  {Buta}, {Paturel}  \& {Fouque}}{{de Vaucouleurs} et~al.}{1991}]{deVauc1991}
{de Vaucouleurs} G.,  {de Vaucouleurs} A.,  {Corwin} Herold~G. J.,  {Buta}
  R.~J.,  {Paturel} G.,   {Fouque} P.,  1991, {Third Reference Catalogue of
  Bright Galaxies}

\bibitem[\protect\citeauthoryear{{van Langevelde}, {Pihlstr{\"o}m}, {Conway},
  {Jaffe}  \& {Schilizzi}}{{van Langevelde} et~al.}{2000}]{vanLangevelde2000}
{van Langevelde} H.~J.,  {Pihlstr{\"o}m} Y.~M.,  {Conway} J.~E.,  {Jaffe} W.,
  {Schilizzi} R.~T.,  2000, \mn@doi [\aap] {10.48550/arXiv.astro-ph/9912437},
  \href {https://ui.adsabs.harvard.edu/abs/2000A&A...354L..45V} {354, L45}

\bibitem[\protect\citeauthoryear{van~der Walt, Colbert  \& Varoquaux}{van~der
  Walt et~al.}{2011}]{walt_numpy_2011}
van~der Walt S.,  Colbert S.~C.,   Varoquaux G.,  2011, \mn@doi [Comput. Sci.
  Eng.] {10.1109/MCSE.2011.37}, 13, 22

\makeatother
\end{thebibliography}

\appendix
\section{Individual source properties}
\label{sec:Appendix}
Below we summarize some of the already known properties of our sample and discuss the molecular emission and absorption in the context of these properties. In many cases we reference molecular masses calculated in section \ref{sec:masses}.

\subsection{NGC6868}
\subsubsection{Galaxy and environment}
NGC6868 is a nearby elliptical and the brightest cluster galaxy of Abell S851. The central radio source has a flat spectrum with a core flux density of 105 mJy at 5 GHz \citep{Hogan2015b}. \citet{Rickes2008} find LINER emission from the galaxy centre due a combination of photoionization, a low-luminosity AGN, and shocks. Recently, \citet{Benedetti2023} have mapped the galaxy's stellar populations using Gemini GMOS IFU. They find it to be dominated by old metal-rich stars, and suggest a historic merger as an explanation for spatially varying metallicity gradients. However, measurements by \citet{Parikh2024} suggest similar abundance gradients may be common and are not necessarily due to mergers.

\subsubsection{Molecular emission}
We find that NGC 6868 has $1.3\pm0.2\times10^{8}\textnormal{M}_{\odot}$ of molecular gas, predominantly contained in a 0.5\,kpc wide and slightly inclined disky structure. There is a very smooth velocity gradient, with the gas at one edge blueshifted to $-120$\,km/s and the other edge redshifted to $210$\,km/s. The velocity dispersion is low at the extremes of the emission but is fairly high closer to the centre, reaching 70\,km/s. There is also a wide tail of molecular emission on the blueshifted wing of the disk -- roughly half as wide as the disky emission. This is of relatively low mass, but is nevertheless clearly detected. Energetic explanations for its origin such as a jet seem unlikely due to the very low velocity dispersion. The merger, suggested by \citep{Benedetti2023} based on the galaxy's spatially varying metallicity gradients, may be a more plausible explanation. If this is indeed the case, the close velocity match between the tail and disky structure suggests the gas has settled. Therefore, this purported merger likely occurred a significant time in the past.

\subsubsection{Molecular and atomic absorption}
CO(1-0) absorption was first detected against the radio core of NGC 6868 by \citet{Rose2019b}. In CO(2-1), CN(2-1), HCN(2-1) and HCO$^{+}$(2-1) we can now see the same absorption regions with higher sensitivity. The two groups of absorption features detected are close to the systemic velocity and appear to be fairly narrow. However, they are also composed of a set of several even narrower lines. The newly detected absorption profiles also have a very high peak optical depth, with the continuum being almost completely absorbed.

\subsection{Abell S555} 

\subsubsection{Galaxy and environment}
The brightest cluster galaxy of Abell S555 lies at the centre of a poorly studied low X-ray luminosity cluster \citep{Bohringer2004}. Its strong central radio source is core dominated, and has a significant gamma-ray flux density. Powerful recent activity has been inferred in the galaxy centre due to its radio spectral energy distribution \citep{Hogan2015b}.

\subsubsection{Molecular emission}
We find that Abell S555 contains $5.6\pm0.5\times10^{8}\textnormal{M}_{\odot}$ of molecular gas, distributed along a line of sight length of around 2.5\,kpc, with a maximum width of around 0.7\,kpc. It has a smooth velocity structure, with the extremes being blueshifted relative to the radio core by \mbox{-100\,km/s} and \mbox{-370\,km/s}. The emission is highly asymmetric and on the most blueshifted side is many times stronger. Interestingly, the more blueshifted side also has a low mass clump of emission at one edge, with a velocity matching that of the least blueshifted emission from the opposite side. Overall, the molecular emission has a bulk velocity -190\,km/s offset from the stellar lines of the radio core, as measured by MUSE. Much of this emission is visible in the CO(1-0) spectrum of Figure \ref{fig:S555_maps_and_spectra}. This is due to its relatively poor angular resolution, which only marginally resolves the emission. In CO(2-1), which has almost five times higher angular resolution, we can see that the continuum source is not spatially coincident with any of the molecular gas. The continuum source and supermassive black hole are therefore offset from the molecular gas in terms of both location and velocity. This indicates a past gravitational disturbance which has led to ram pressure forces stripping the AGN of its surrounding molecular gas.

\subsubsection{Molecular and atomic absorption}
H\thinspace\small I\normalsize\space absorption has been searched for against the radio core of Abell S555, providing an upper limit of $\tau_{\textnormal{max}}<0.013$ \citep{Hogan_thesis}. A single CO(1-0) absorption feature was first detected in Abell S555 by \citet{Rose2019b}. With our new observations we are able to re-detect this absorption, plus one new and slightly less redshifted feature at +190\,km/s. Compared with the absorption detected in other systems, these lines are moderately wide. They are also $>400$\,km/s offset from any of the molecular emission, as well as being spatially offset. This suggests the absorbing clouds cannot be a part of the same population which are visible via the molecular emission.

\subsection{Hydra-A}

\subsubsection{Galaxy and environment}
Hydra-A is a giant elliptical galaxy with an edge-on disk of dust and molecular gas which lies at the centre of an X-ray bright cluster \citep{Hamer2014}. At 200\,GHz, powerful radio lobes propagate 40\,kpc to the north and south from the centre \citep{Taylor1990}, but at 1\,GHz these extend to over 500\,kpc in the north and 250\,kpc in the south \citep{Lane2004, Baidoo2023}. The radio lobes are surrounded by cavities in the X-ray emitting gas of the intracluster medium created by repeated AGN outbursts \citep{McNamara2000, Wise2006}.

\subsubsection{Molecular emission}
Hydra-A has a $3.1\pm0.2\times10^{9}\textnormal{M}_{\odot}$ edge-on molecular disk with a very smooth velocity gradient and a central velocity matching that of of the stellar absorption lines of the radio core \citep{Rose2019a}. Most of the disk has a velocity dispersion in the range of 30-50\,km/s, and the two extreme edges are rotating at close to -400 and +400\,km/s. 

\subsubsection{Molecular and atomic absorption}
Hydra-A has by far the best studied intrinsic absorption detected in a brightest cluster galaxy. A total of 12 molecular clouds or groups of clouds with velocity dispersions as low as 0.7 km/s have been detected via 11 different molecular species \citep{Rose2020}. H\thinspace\small I\normalsize\space absorption has also been detected against the core of the galaxy with a peak optical depth of $\tau$~=~0.0015 \citep{Taylor1996}. Five of the tracers in which absorption has been detected are illustrated in Figure \ref{fig:HydraA_maps_and_spectra}, and the rest can be seen in \citet{Rose2020}. Interestingly, the velocities at which the absorption is found match those of the clouds observed through molecular emission. The absorption lines may therefore represent a subsample of the clouds visible in the disk's emission \citep{Rose2020}.

\subsection{Abell 2597}
\subsubsection{Galaxy and environment}
The brightest cluster galaxy of Abell 2597 is a giant elliptical at the centre of a cool core. It is hosted by dense halo of hot and bright X-ray emitting plasma of megaparsec scales, within which are buoyantly rising cavities inflated by the radio jets and lobes of the central galaxy \citep{McNamara2001}. The gas cooling towards the central galaxy is fueling active star formation at a rate of $5\textnormal{M}_{\odot}/$year \citep{Tremblay2012}.

\subsubsection{Molecular emission}
Abell 2597 is extremely gas rich, with a molecular mass in its central regions of $4.6\pm0.5\times10^{9}\textnormal{M}_{\odot}$ (by central regions, we refer the the approximate region shown in Fig \ref{fig:A2597_maps_and_spectra}). However, there are also filaments of molecular gas to the north and south extending out to around 20\,kpc \citep{Tremblay2016, Tremblay2018}.

\subsubsection{Molecular and atomic absorption}
Three CO(2-1) absorption lines have been detected against the bright and unresolved radio core \citep{Tremblay2016}. These have optical depths of $\tau \sim 0.2-0.3$ and velocities of $240-335$\,km/s, with their positive sign implying movement towards the galaxy centre along the line of sight. In HCN(2-1) and HCO$^{+}$(2-1), we also detect some weaker absorption at velocities of around 120-200\,km/s. In CO(2-1), all of the absorption features are fairly wide. However, in HCN(2-1) and HCO$^{+}$(2-1), they become progressively wider and are blended together. In HI, they are much wider still and form into a single absorption feature almost 1000\,km/s wide. \citet{Tremblay2018}, suggest that the CO(2-1) emission and absorption do not overlap in terms of their velocity. However, the additional absorption we have found at 120-200\,km/s does appear to match the more extreme redshifted edges of the emission. We also find some additional CO(2-1) absorption which overlaps with the emission.

\subsection{Abell 2390}
\subsubsection{Galaxy and environment}
The brightest cluster galaxy in Abell 2390 lies in an X-ray luminous galaxy cluster with a cooling rate of $200-300\,\textnormal{M}_{\odot}\textnormal{yr}^{-1}$ \citep[][]{Allen2001}. The central continuum source has a flat light curve dating back to at least 2013, but recent activity is indicated by high angular resolution observations \citep{Augusto2006,Rose2020}. These show young, compact and self-absorbed jets, with half the flux coming from synchrotron emission and half from dust at 850\,$\mu$m \citep{Edge1999, Augusto2006}. Significantly older radio activity is indicated by X-ray cavities in the cluster's hot atmosphere, as well as $300-600$\,kpc wide radio lobes \citep{Savini2019}. 

\subsubsection{Molecular emission}
Abell 2390 has a highly asymmetric plume of multiphase gas, which extends out to 15\,kpc in CO and 25\,kpc in H$\alpha$ \citep[]{Alcorn2023, Rose2024}. This plume may have been driven by a jet and aided by jet inflated X-ray bubbles, or be a trail of gas stripped from the galaxy following a gravitational disturbance \citep{Rose2024}. 

\subsubsection{Molecular and atomic absorption}
H\thinspace\small I\normalsize\space absorption has been detected against the radio core with $\tau_{\textnormal{max}}=0.084 \pm 0.011$ \citep{Hernandez2008, Hogan_thesis}. CO(1-0) absorption was first detected by \citep{Rose2019b}, to which we now add CO(2-1), HCN(2-1) and HCO$^{+}$(2-1) absorption. Across all molecules, the absorption profile is extremely broad and contains two overlapping components. There are no hints of any narrow sub-features, as in e.g. Centaurus-A and Hydra-A. As well as being broad, the absorption is at or close to saturation in HCN(2-1) and HCO$^{+}$(2-1). Although the absorption is entirely redshifted, we also see CO(2-1) emission with matching velocities along the line of sight to the radio core. However, this may be due to the observation's wide beam encapsulating emission from far beyond the radio core.

\subsection{RXCJ0439.0+0520}
\subsubsection{Galaxy and environment}
The brightest cluster galaxy of RXCJ0439.0+0520 is not as well studied as most of this sample, but is known to have a high H$\alpha$ luminosity of \mbox{$6 \times 10^{40}$ erg s$^{-1}$} detected by the VIsible Multi Object Spectrograph (VIMOS) of the Very Large Telescope (VLT) \citep{Hamer2016}. The central continuum source's light curve has also shown significant variability in the past, but has been relatively constant between 2010 and 2022 \citep{Hogan2015, Rose2022}.

\subsubsection{Molecular emission}
In RXCJ0439.0+0520 we find $1.3\pm0.4\times10^{10}\textnormal{M}_{\odot}$ of molecular gas, the bulk of which is in a marginally resolved clump centred on the radio core. The remaining gas lies along a 2\,kpc wide and 10\,kpc long plume. Much of this plume is only weakly detected and it is hard to constrain its velocity. However, a central bright region in the plume appears to have a very similar velocity to the gas centred on the continuum source. This implies the elongation of the plume is almost perpendicular to our line of sight. Without $\sim$\,GHz frequency radio observations, it is not possible to determine whether or not previous radio activity may have been responsible for the plume. However, the low velocity dispersion suggests this explanation is unlikely. As in Abell 2390, ram pressure forces due to a past gravitational disturbance are the most plausible alternative explanation. 

\subsubsection{Molecular and atomic absorption}
CO(1-0) and CN(1-0) absorption has previously been detected against the continuum by \citet{Rose2019b}, but H\thinspace\small I\normalsize\space absorption has not been found \citep[with an upper limit of $\tau_{\textnormal{max}}<0.13$][]{Hogan_thesis}. The molecular absorption previously detected is wide and very close the the systemic velocity. However, with CO(2-1), CN(2-1), HCN(2-1) and HCO$^{+}$(2-1), we now detect a second absorption feature. This is similarly wide, but is redshifted to 550\,km/s. Interestingly, this is more strongly detected than the 0\,km/s feature in all of these molecules, despite its absence from the CO(1-0) spectrum. The emission coincident with the radio core is only weakly detected, but has a large velocity width. This overlaps with the 0\,km/s feature, but due to the noise of the spectrum it is difficult to be certain whether or not it overlaps with the 550\,km/s feature.

\subsection{Abell 1644}
\subsubsection{Galaxy and environment}
The brightest cluster galaxy of Abell 1644 is a complex system at the centre of the brighter of two X-ray peaks in its gravitationally disturbed host cluster \citep{Johnson2010}. In recent years the central radio source has been highly variable, with its 15\,GHz flux increasing from around 160\,mJy in early 2018 to around 240\,mJy in early 2021 \citep{Rose2022}.

\subsubsection{Molecular emission}
Abell 1644's molecular emission has previously been detected in CO(1-0), but in CO(2-1) we are now able to see it with much higher angular resolution and sensitivity. Based on CO(1-0) emission, \citet{Baek2022} identify clumps of molecular gas in an arc along the sloshing patterns detected in X-ray emission. With our CO(2-1) observations, we are able to re-detect all but the farthest of these clumps at higher angular resolution. As such we now deconvolve what appeared to be a single, rotating structure centred close to the continuum source. The CO(2-1) maps in Figure \ref{fig:A1644_maps_and_spectra} now reveal a dearth of molecular gas coincident with the continuum. Instead, the gas is separated into multiple components 1 -- 2 kpc from the continuum. One component is compact, has a comparatively high velocity dispersion and is marginally redshifted relative to the stellar absorption lines of the core. Another is disky, approximately 4\,kpc long, has a central velocity of -140\,km/s and a velocity change of 100\,km/s. The velocities of these two central components are more extreme than the red and blueshifted clumps which lie several kpc further from the radio continuum source. They are also separated by a third, weaker region of molecular emission.

\subsubsection{Molecular absorption}
CO(1-0) and CN(1-0) absorption has been detected against the central radio continuum source \citep{Rose2019b, Baek2022}. Our new observations of CO(2-1), CN(2-1), CS(5-4), HCN(2-1) and HCO$^{+}$(2-1) detect the absorption to much higher significance and show multiple new features. Some of these are sub-components of the absorption previously detected, while others are at neighbouring velocities. Generally these absorption components are of moderate width and centred at the systemic velocity. However, we also find a weak but very wide absorption feature centred at 150\,km/s. Our high sensitivity and high angular resolution CO(2-1) observations show that there is little or no molecular gas along the line of sight to the continuum source. We can therefore be confident that the clouds responsible for the absorption in Abell 1644 are not a subsample of those which we see in emission elsewhere in the galaxy.

\subsection{Circinus}
\subsubsection{Galaxy and environment}
Circinus is one of the closest known active galaxies, at a distance of $4.2\pm0.7$\,Mpc and a redshift of 0.0015 \citep{Tully2009}. In the X-ray, it is the brightest known Compton-thick Seyfert 2 galaxy \citep{Bianchi2002}. On one side, Circinus has kiloparsec scale outflowing gas seen in H$\alpha$ and [O III] emission lines \citep{Veilleux1997,MullerSanchez2006}. In the central 100\,pc, \citet{Esquej2014} find a star formation rate of approximately 0.1\,M$_{\odot}$/yr, with the total for the galaxy estimated at a few M$_{\odot}$/yr \citep{Elmouttie1998}.

\subsubsection{Molecular emission}
\citet{Izumi2023} have presented HCN(3-2) emission which reveals a roughly 3\,pc wide circumnuclear disk. On a larger scale and width CO(3-2), Circinus is found to contain spiral arms rich in molecular gas which are inclined at an angle of $\approx 65^{\circ}$, but they do not pass through our line of sight to the continuum. Absorption is therefore unlikely to arise from this large scale distribution of gas. 

\subsubsection{Molecular absorption}
Circinus contains CO(3-2), HCN(4-3), HCO$^{+}$(3-2) and HCO$^{+}$(4-3) absorption \citep{Izumi2023}. Although the HCO$^{+}$(4-3) absorption is near saturated, the HCO$^{+}$(3-2) spectrum is the most sensitive. Here, two absorption features are present at 5 and 60\,km/s. We also identify a third, weaker absorption feature at 420\,km/s. Each of the molecular species in Figure \ref{fig:Circinus_maps_and_spectra} also contains strong emission, suggesting the gas of the circumnuclear disk is very dense. Excluding the 420\,km/s feature, the velocity of the absorbing gas is at or close to the peak of the emission of the wider circumnuclear disk in each species.

\subsection{NGC 4261}
\subsubsection{Galaxy and environment}
NGC 4261 is an elliptical in the Virgo cluster with among the brightest radio fluxes of low redshift \citep[z$<0.01$,][]{Gavazzi2004} FR\,I radio galaxies ($\sim$\,19\,Jy at 1.4 GHz, $\sim$\,205\,mJy at 300\,GHz). The axis of its radio jets is closely aligned with a 200\,pc wide disky structure of dust and molecular gas traced with CO, HCN and HCO$^{+}$ \citep{Jaffe1994, Boizelle2021, Swada-Satoh2022, Ruffa2023}. In the innermost regions, H$_{2}$O megamaser emission is reported by \citet{Swada-Satoh2022}. \citet{Ruffa2023} determine the presence of a $(1.62\pm0.04)\times10^{9}\textnormal{M}_{\odot}$ supermassive black hole.

\subsubsection{Molecular emission}
From CO(2-1) emission, we estimate a molecular mass of $2.1\times 10^{7}\textnormal{M}_{\odot}$. Accurate dimensions of NGC 4261's molecular disk are harder to determine than for the dust disk \citep{Ferrarese1996}, so its inclination is uncertain. In the CO(2-1) flux map (Figure \ref{fig:NGC4261_maps_and_spectra}), the disc is roughly circular, implying it is close to face-on. However, with $>4\sigma$ emission (see its velocity map), the disk appears more elongated. Nevertheless, its width of only 150\,pc means it cannot be considered a galaxy scale disk

\subsubsection{Molecular and atomic absorption}
Absorption features redshifted by 60\,km/s have been seen against the central continuum in CO(1-0) and HI \citet{Jaffe1994, vanLangevelde2000}. From archival observations, we also find serendipitous detections of HCN(4-3) and HCO$^{+}$(4-3) absorption. The spectra are noisy, but appear to contain two slightly redshifted and overlapping absorption features.

\subsection{NGC 5044}
\subsubsection{Galaxy and environment}
NGC 5044 is an elliptical brightest group galaxy at the centre of a moderate cooling flow, and hosts a substantial mass of multiphase gas \citep{David2011}. Surrounding the galaxy are numerous cavities and X-ray filaments which have been inflated by the AGN \citep{Buote2004,David2011,Gastaldello2013}. Among galaxy groups, NGC 5044's has the brightest known H$\alpha+$[NII] filaments \citep{David2014}.

\subsubsection{Molecular emission}
CO(2-1) observations imply a molecular gas mass of $4 - 6 \times 10^{7}~\textnormal{M}_{\odot}$, off-centre from the continuum source \citep{Temi2018, Schellenberger2020}. Much higher angular resolution CO(1-0) observations presented by \citet{Rose2023} appear to show a lack of molecular gas along the line of sight to the radio core. Instead, the gas appears to be within one or two clumps approximately 0.5\,kpc away. 

\subsubsection{Molecular absorption}
CO(2-1) absorption in NGC 5044 was first reported by \citet{David2014}. We are able to re-detect the same absorption feature in HCN(2-1) and HCO$^{+}$(2-1). The absorption is much wider than is detected in many of the disky sources such as Centaurus-A and Hydra-A, but is also narrower than in some systems such as Abell 2390 and Abell 2597. None of the galaxy's molecular emission is spatially coincident with the absorption, suggesting they are not caused by the same cloud population.

\subsection{NGC 1052} 
\subsubsection{Galaxy and environment}
NGC1052 contains a low-luminosity AGN with faint X-ray emission \citep{Woo2002}. Despite the AGN's low luminosity, it has produced radio jets on pc to kpc scales which suggest recent and historic accretion \citep{Nakahara2020}. \citet{Kameno2024} have reported sub-pc scale observations of 321\,GHz H$_2$O emission, indicating water masers incredibly close to the supermassive black hole's event horizon, but with remarkably similar dynamics to the molecular absorption.

\subsubsection{Molecular emission}
High angular resolution ALMA observations presented by \citet{Kameno2020} show CO emission from a rotating ring-like circumnuclear disk with a molecular mass of $5.3\times 10^{5}~\textnormal{M}_{\odot}$. 

\subsubsection{Molecular absorption}
\citet{Kameno2020} also show a rich set of absorption features detected against the radio continuum source, including CO, HCN, HCO$^{+}$, SO, SO$_{2}$, CS and CN. Wide absorption lines are present from approximately -200 to 200\,km/s, with the redshifted absorption being stronger.

\subsection{IC 4296} 
\subsubsection{Galaxy and environment}
IC 4296 is an elliptical galaxy and the brightest cluster galaxy of Abell 3565. It hosts the FRI radio source PKS 1333-33, contains a nuclear dust disk and has a large-scale, double-sided jet that extends up to 77 kpc from the nucleus \citep{Lauer2005, Burke2009}. The inner jet is unresolved at 230\,GHz but faint emission is detected at a distance of 950 pc North-West of the nucleus \citep{Ruffa2019}. 

\subsubsection{Molecular emission}
IC 4296 has disky CO(2-1) emission with a mass of 2.3$\pm0.2\times10^{7}\textnormal{M}_{\odot}$. The velocity field shows an s-shaped zero-velocity contour, suggesting the presence of a warp in the disk \citep{Ruffa2019}. 

\subsubsection{Molecular absorption}
Against the radio core there is a narrow CO(2-1) absorption feature with a velocity closely matching that of the galaxy and its molecular emission \citep{Ruffa2019}.

\bsp	
\label{lastpage}
\end{document}